\documentstyle[aps,prd,preprint]{revtex}

\tighten
\begin{document}
\draft

\preprint{\vbox{\baselineskip=12pt
\rightline{gr-qc/9503025}
\rightline{WUGRAV-95-4}
\rightline{Submitted to Physical Review D}}}

\title{Three dimensional numerical relativity:  the evolution of black holes}

\author{Peter Anninos${}^{(1)}$, Karen Camarda${}^{(1,4)}$, Joan
Mass\'o${}^{(1,2)}$, Edward Seidel${}^{(1,4)}$, Wai-Mo Suen${}^{(3)}$,
and John Towns${}^{(1)}$
}

\address{
${}^{(1)}$ National Center for Supercomputing Applications \\
605 E. Springfield Ave., Champaign, Il. 61820
}

\address{
${}^{(2)}$ Department of Physics \\
Universitat de les Illes Balears, Palma de Mallorca E-07071, Spain
}

\address{
${}^{(3)}$ Department of Physics \\
Washington University, St. Louis, MO 63130
}

\address{
${}^{(4)}$ Department of Physics \\
University of Illinois at Urbana-Champaign, Urbana, IL 61801
}
\maketitle

\begin{abstract}

We report on a new 3D numerical code designed to solve the Einstein
equations for general vacuum spacetimes.  This code is based on the
standard 3+1 approach using Cartesian coordinates.  We discuss the
numerical techniques used in developing this code, and its performance
on massively parallel and vector supercomputers. As a test case, we
present evolutions for the first 3D black hole spacetimes.  We
identify a number of difficulties in evolving 3D black holes and
suggest approaches to overcome them.  We show how special treatment of
the conformal factor can lead to more accurate evolution, and discuss
techniques we developed to handle black hole spacetimes in the absence
of symmetries.  Many different slicing conditions are tested,
including geodesic, maximal, and various algebraic conditions on the
lapse.  With current resolutions, limited by computer memory sizes, we
show that with certain lapse conditions we can evolve the black hole
to about $t=50M$, where $M$ is the black hole mass.  Comparisons are
made with results obtained by evolving spherical initial black hole
data sets with a 1D spherically symmetric code.  We also demonstrate
that an ``apparent horizon locking shift'' can be used to prevent the
development of large gradients in the metric functions that result
from singularity avoiding time slicings.  We compute the mass of the
apparent horizon in these spacetimes, and find that in many cases it
can be conserved to within about 5\% throughout the evolution with our
techniques and current resolution.

\end{abstract}


\widetext

\section{introduction}
\label{sec:introduction}

Progress in three dimensional numerical relativity has been impeded in
part by a lack of computers with sufficient memory and computational
power to perform well resolved calculations of 3D spacetimes.  To
date, only a few groups have attempted full 3D numerical relativity
calculations, notably the Kyoto and Texas groups.  The Kyoto group has
applied themselves to pure gravitational wave spacetimes in the
linearized limit~\cite{Nakamura87}, and more recently to general
relativistic hydrodynamical studies of revolving neutron stars.  The
Texas group has designed a code for the simulation of various
cosmological spacetimes~\cite{Laguna91}, and has recently turned to
the problem of black hole spacetimes~\cite{Laguna93}.  Such 3D
calculations in numerical relativity have proved quite difficult due
to constraints placed by the largest computers available.

However, these artificial restrictions on the physical simulations
dictated by memory and speed considerations are being relaxed
considerably due to the introduction of massively parallel machines.
Machines available today have gigabytes of memory and are capable of
speeds of tens of gigaflops, allowing a completely new class of
problems to be investigated. It is our intention to develop general
purpose numerical codes to model general 3D spacetimes, including
dynamic multiple black hole spacetimes. Simulations of general black
hole interactions are inherently three dimensional problems.  To date,
numerical simulations of black hole spacetimes have been limited to
two dimensional axisymmetric geometries
\cite{Anninos93c,Anninos93b,Bernstein93b}. A Grand Challenge effort is
currently under way in the numerical relativity community to develop
three dimensional codes that will be applied, for example, to the
coalescence of binary black hole systems. Some progress has already
been made to construct initial data in three dimensions representing
two black hole configurations with arbitrary positions, radii, linear
momenta and spin \cite{Cook93}.

In this paper we report on progress made in the development of a new
3D numerical code based on the standard ADM or 3+1 approach
\cite{Arnowitt62} that we call the ``G'' code.  We have also developed
another 3D code based on a completely different formulation of the
Einstein equations due to Bona and Mass\'o \cite{Bona92,Bona93} that
we call the ``H'' code.  This promising formulation and a recent
extension \cite{Bona94b} casts the equations in a first order, flux
conservative, hyperbolic (FOFCH) form that allows very accurate and
sophisticated numerical methods to be applied to the Einstein
equations for the first time.  Applications of this formalism to
black holes and gravitational waves are very promising in 1D and 3D
studies.  We will report on results from the FOFCH codes in a future
paper.

Here we apply the ADM code to the problem of black hole spacetimes,
showing the first evolutions of black holes in a 3D Cartesian
coordinate system.  This is a difficult problem in 3D due to the large
gradients that typically develop near the black hole, and we report on
progress we have achieved towards evolving a spherical black hole to
about $50M$ in time, where $M$ is the mass of the black hole.  This
time scale is many characteristic time scales of the black hole, as
the light crossing time of the black hole is $4M$, while its
fundamental quasinormal mode period is about $17M$.

There are several reasons why we study the spherical black hole
spacetime before looking at more general spacetimes.  First, this
system has been studied extensively in 1D and
2D~\cite{Anninos93c,Bernstein89,Seidel92a,Abrahams92a}, so it can be
used as a standard system for testing out numerical schemes.  Second,
evolving a spherical black hole is a difficult exercise, as shown in
Refs.~\cite{Bernstein89,Seidel92a}, even when the coordinate system is
chosen to match the underlying geometry of the system (e.g.,
spherical-polar coordinates).  It represents one of the first major
challenges to 3D black hole evolutions since it is the longitudinal,
or spherical part of the calculations that presents the most serious
problems in terms of the gradients in metric functions that develop as
the system evolves.  Since we are solving this problem using 3D {\em
Cartesian} coordinates, the spherical system has no special symmetry
from the point of view of the code.  By concentrating first on solving
the problems associated with evolving this system, we will learn the
techniques that are required to evolve much more general black hole
spacetimes. Finally, the Schwarzschild spacetime is the endpoint for
any generic 3D black hole spacetime that does not have net angular
momentum, so it is important to understand the problems associated
with the end state of more complex calculations that will be performed
in the future.

We should point out that there is work going on in our group using the
same 3D relativity code to evolve gravitational waves. This work will
be described in a separate paper \cite{Anninos94d}.  Our strategy is
to study first the issues of black holes without the complications of
gravitational waves, and waves without the complications of black
holes, and then to combine them when we consider distorted black holes
that will evolve and emit gravitational waves as they ``ring down'' to
Schwarzschild (see, e.g., Ref.~\cite{Abrahams92a}).

We discuss the theoretical foundations of the ADM approach and the
numerical algorithms used to solve the resulting equations in sections
\ref{sec:math} and \ref{sec:numeric}, respectively.  The initial data
is discussed in section \ref{sec:initial} and various issues regarding
black hole evolutions in 3D, including computational approaches and
gauge conditions are discussed elsewhere in section
\ref{sec:evolution}.  In section \ref{sec:results} we present case
study tests of different slicings and shift conditions.  Finally in
section \ref{sec:summary} we discuss the results and future directions
for this work.  Appendix~\ref{sec:appendix} provides information about
programming and performance issues encountered in developing a
massively parallel 3D code.

\section{Mathematical Development}
\label{sec:math}

We use the standard 3+1 ADM approach \cite{York79} to write the
general spacetime metric in the form
\begin{equation}
 ds^2   = -(\alpha^{2} -\beta ^{a}\beta _{a}) dt^2
 + 2 \beta _{a}  dx^{a} dt
 +\gamma_{ab} dx^{a} dx^{b},
\end{equation}
using geometrized units such that the gravitational constant $G$ and
the speed of light $c$ are both equal to unity. Throughout this paper,
we use Latin indices to label spatial coordinates, running from 1 to
3. The lapse function
$\alpha$ and the shift vector $\beta^{a}$ determine how the slices are
threaded by the spatial coordinates.  Together, $\alpha$ and $\beta^a$
represent the coordinate degrees of freedom inherent in the covariant
formulation of Einstein's equations, and can therefore be chosen
freely.  Various choices used in our code are discussed in
sections~\ref{sec:lapse} and~\ref{sec:shift}.

The Ricci tensor of the spacetime may be decomposed into its spatial
and timelike components, and when the vacuum Einstein equations are
imposed these reduce to the four constraint equations
\begin{equation}
\label{Hamiltonian constraint}
R+({\mathrm tr} K)^{2}-K^{ab}K_{ab}=0,
\end{equation}
\begin{equation}
\label{momentum constraint}
D_{b}(K^{ab}-\gamma^{ab}{\mathrm tr} K)=0,
\end{equation}
and the twelve evolution equations
\begin{equation}
\label{metric evolution}
\partial_{t}\gamma_{ab}=-2\alpha K_{ab}+
D_{a}\beta_{b}+D_{b}\beta_{a},
\end{equation}
\begin{equation}
\label{excurv evolution}
\partial_{t}K_{ab}=-D_{a}D_{b}\alpha+\alpha \left[
R_{ab}+\mbox({tr}K)K_{ab}-2K_{ac}K^{c}{}_{b} \right]
+\beta^{c}D_{c}K_{ab}+
K_{ac}D_{b}\beta^{c}+K_{cb}D_{a}\beta^{c}.
\end{equation}
Here $R_{ab}$ is the Ricci tensor, $R$ the scalar curvature, and
$D_{a}$ the covariant derivative associated with three-dimensional
metric $\gamma_{ab}$.  The Einstein equations are contained in Eqs.
(\ref{Hamiltonian constraint}), (\ref{momentum constraint}) and
(\ref{excurv evolution}), while Eq. (\ref{metric evolution}) follows
from the definition of the extrinsic curvature $K_{ab}$.  In our work
the constraints are solved on an ``initial'' hypersurface using the
well known conformal decomposition method of York and
coworkers~\cite{York79} and then evolved forward in time using the
evolution equations (\ref{metric evolution}) and (\ref{excurv
evolution}).

If the constraints are satisfied on any hypersurface, the Bianchi
identities then guarantee that they remain satisfied on all subsequent
hypersurfaces.  In a numerical solution, this may not be the case and
the constraints have to be monitored carefully in order to ensure that
the spacetimes generated are accurate.  Traditional alternatives to
this approach involve solving the constraint equations on each slice
for certain metric and extrinsic curvature components, and then simply
monitoring the ``left over'' evolution equations.  This issue is
discussed further by Choptuik in Ref.~\cite{Choptuik91}, and in detail
for the Schwarzschild spacetime in Ref.~\cite{Bernstein89}.  New
approaches to this problem of constraint vs.  evolution equations are
currently being pursued~\cite{Lee93,Lee94a}.

\section{Numerical Algorithms}
\label{sec:numeric}

In this section we discuss the various numerical algorithms we have
developed to evolve the spacetime.  The methods presented here are not
specialized to black holes but apply generally to all systems we plan
to study with this code.  We have developed methods for solving both
the hyperbolic evolution equations and various elliptic equations,
such as the initial data and maximal slicing equations.  The numerical
grid we use is a fixed Cartesian grid with constant spacing between
spatial grid points.  Cartesian grids have the advantage of covering
the spacetime with coordinates that are inherently singularity free.
To date we have considered only fixed, constant time step sizes in our
evolution schemes, discussed below.  We utilize the conventional
indices $i$, $j$, and $k$ to label the space steps, and $n$ to label
the time steps in the finite differenced forms of the evolution
equations.

\subsection{Hyperbolic Equations}
\label{sec:hyperbolic}

The use of MACSYMA scripts written by David Hobill \cite{Anninos93c}
to generate symbolic expressions for the ADM form of the Einstein
equations and the ability of MACSYMA to translate symbolic expressions
into FORTRAN code tremendously reduced the amount of time spent on
laborious calculations and allowed us to write a 3D code in a short
period of time.  Our numerical methods are adopted from previous work
in developing 2D axisymmetric codes. We briefly describe those methods
here but refer the reader to references
\cite{Anninos93c,Anninos93b,Bernstein93b} for further details.

The explicit leapfrog method is used to evolve the hyperbolic system
of equations explicitly in time to second order accuracy.  In the
conventional leapfrog method, the extrinsic curvature ($K_{ab}$)
variables are offset a half step in time relative to the metric field
variables ($\gamma_{ab}$).  Placing the metric components at the half
timestep and the extrinsic curvature at the full step, we can write
the finite differenced forms of (\ref{metric evolution}) and
(\ref{excurv evolution}) schematically as
\begin{equation}
\gamma_{i,j,k}^{n+1/2}=\gamma_{i,j,k}^{n-1/2} -2\alpha_{i,j,k}^n
                       K_{i,j,k}^n \Delta t
\end{equation}
and
\begin{equation}
K_{i,j,k}^{n+1}=K_{i,j,k}^n+\left\{\alpha_{i,j,k}^n\left[
                \left(K_{i,j,k}^n\right)^2+R_{i,j,k}^{n+1/2}\right]
          -\left(\nabla^2\alpha\right)_{i,j,k}^n\right\}\Delta t,
\end{equation}
where we have dropped the tensor index notation to prevent confusion with
the indices $i$, $j$, and $k$ which are used to locate a quantity on
the spatial grid, and where a number of terms have been omitted for
clarity.

These equations introduce first order errors due to the placement of
the lapse function and the nonlinear terms $(K_{i,j,k}^n)^2$. We
adjust for this problem by extrapolating the necessary variables to
the $n+\frac{1}{2}$ time slice using the formula
\begin{equation}
K_{i,j,k}^{n+1/2} =\frac{3}{2} K_{i,j,k}^n -\frac{1}{2} K_{i,j,k}^{n-1}.
\end{equation}
We have used this method of solution in previous work
\cite{Anninos93c,Anninos93b,Bernstein93b} and found it to work
very well.

Spatial first and second derivatives of the metric components present
in the terms $(\nabla\nabla \alpha)_{i,j,k}$ and $R_{i,j,k}$ are
defined using either standard second or fourth order center
differences.  The added complexity of higher order differences poses
certain performance problems with regard to parallel machines. These
issues are discussed in appendix A.  We found in previous work
\cite{Anninos93c,Anninos93b,Bernstein93b} that fourth order
differences provide more accurate solutions and propagate
gravitational waves with less dispersion and damping than the second
order differences.  Our code is designed to allow for both second and
fourth order differences as options.  However, fourth order
differences are more unstable than second order, particularly at late
times in the evolution when large gradients develop near the horizon.
All results presented in this paper were obtained with second order
derivatives.

\subsection{Elliptic Equations}
\label{sec:elliptic}

The 3D code allows for an arbitrary set of initial conditions, time
slicings, and gauge conditions for the spacetime, all of which can
require solutions to elliptic equations.  For example, in previous
work (see, e.g., \cite{Anninos93c} or \cite{Bernstein93b}) maximal
slicing has frequently been used for black hole spacetimes due to its
singularity avoiding nature and smooth properties (although this is
not necessarily the case in 3D, as we show in section
\ref{sec:maxresults}).  Satisfying the maximal slicing condition
requires the solution of an elliptic equation, however. Furthermore,
the initial data problem is usually formulated in terms of elliptic
equations that must be solved on the initial hypersurface.  The
Schwarzschild initial data evolved in this paper is known
analytically, but for more general initial data sets we often will
need to solve an elliptic PDE.  The solution of elliptic equations,
particularly on large meshes and on parallel computers, is a costly
operation.

At the time this code was developed, there was no general purpose
package for solving linear systems of equations on parallel machines.
Furthermore, we found previously that many standard packages available
on other machines (e.g., Cray) were not efficient enough to solve
elliptic equations in a time that made them effective for these
problems~\cite{Bernstein93b,Towns92}.  This prompted us to develop a
specialized 2D solver for the Cray and then to develop a more general
3D solver called CMStab~\cite{Towns93}, initially for the CM-5, to
provide a variety of Conjugate Gradient and Conjugate Gradient-like
algorithms for solving the elliptic equations.  The CMStab code is
publicly available and can be obtained from our Web server (see
section \ref{sec:summary}). We are presently moving it to other
architectures.

\subsubsection{Generating the linear system}

In order to solve elliptic equations numerically, the usual approach
is to approximate the derivatives with finite difference operators and
solve the resulting system of simultaneous equations.  Fortunately,
our elliptic equations are linear, so the resulting system is a linear
system of simultaneous equations.  (Nonlinearities are typically
handled by linearizing and iterating, so the technique is essentially
the same.) As an illustration of this process, the maximal slicing
equation for the lapse function $\alpha$ will be used (see section
\ref{sec:maximal} below for a discussion of this condition):
\begin{eqnarray}
\label{maximal slicing}
     && \partial_x (
                   \sqrt{\gamma} (
                                   \gamma^{xx} \partial_x \alpha +
                                   \gamma^{xy} \partial_y \alpha +
                                   \gamma^{xz} \partial_z \alpha
                                 )
                 ) + \nonumber \\
     && \partial_y (
                   \sqrt{\gamma} (
                                   \gamma^{xy} \partial_x \alpha +
                                   \gamma^{yy} \partial_y \alpha +
                                   \gamma^{yz} \partial_z \alpha
                                 )
                 ) + \nonumber \\
     && \partial_z (
                   \sqrt{\gamma} (
                                   \gamma^{xz} \partial_x \alpha +
                                   \gamma^{yz} \partial_y \alpha +
                                   \gamma^{zz} \partial_z \alpha
                                 )
                 )
 = \sqrt{\gamma} K^{ab} K_{ab} \alpha.
\end{eqnarray}
Here, $\gamma$ is the determinant of the covariant 3-metric. Eq.
(\ref{maximal slicing}) involves
both first and second order derivatives of $\alpha$.  Second order
accurate, central finite differences are used.  Since the differencing is
done in three dimensions and all mixed derivatives are
involved, we have a nineteen-point stencil leading to the following
equation, valid at each grid point $(i,j,k)$ not on the boundary:
\begin{eqnarray}
\label{fdalp}
&&C^{(1)}_{i,j,k} \alpha_{i,j,k} +
  C^{(2)}_{i,j,k} \alpha_{i+1,j,k} +
  C^{(3)}_{i,j,k} \alpha_{i-1,j,k} +
  C^{(4)}_{i,j,k} \alpha_{i,j+1,k} + \nonumber \\
&&C^{(5)}_{i,j,k} \alpha_{i,j-1,k} +
  C^{(6)}_{i,j,k} \alpha_{i,j,k+1} +
  C^{(7)}_{i,j,k} \alpha_{i,j,k-1} +
  C^{(8)}_{i,j,k} \alpha_{i+1,j+1,k} + \nonumber \\
&&C^{(9)}_{i,j,k} \alpha_{i+1,j-1,k} +
  C^{(10)}_{i,j,k} \alpha_{i-1,j+1,k} +
  C^{(11)}_{i,j,k} \alpha_{i-1,j-1,k} +
  C^{(12)}_{i,j,k} \alpha_{i+1,j,k+1} + \nonumber \\
&&C^{(13)}_{i,j,k} \alpha_{i+1,j,k-1} +
  C^{(14)}_{i,j,k} \alpha_{i-1,j,k+1} +
  C^{(15)}_{i,j,k} \alpha_{i-1,j,k-1} +
  C^{(16)}_{i,j,k} \alpha_{i,j+1,k+1} + \nonumber \\
&&C^{(17)}_{i,j,k} \alpha_{i,j+1,k-1} +
  C^{(18)}_{i,j,k} \alpha_{i,j-1,k+1} +
  C^{(19)}_{i,j,k} \alpha_{i,j-1,k-1} = 0.
\end{eqnarray}
The stencil coefficients $C^{(n)}$ are obtained by finite differencing the left
hand side of the maximal slicing equation in such a way as to make the
resulting matrix symmetric, at least before the application of
boundary conditions. For terms of the form $\partial_x (f \partial_x
\alpha)$, a finite difference equation that results in a symmetric
matrix is
\begin{eqnarray}
\partial_x (f \partial_x \alpha) = \frac{1}{2 \Delta x^2}
      (\alpha_{i+1,j,k} (f_{i+1,j,k}+f_{i,j,k}) -
       &&\alpha_{i,j,k} (f_{i+1,j,k}+2f_{i,j,k}+f_{i-1,j,k}) +
\nonumber \\
       &&\alpha_{i-1,j,k} (f_{i,j,k}+f_{i-1,j,k})),
\end{eqnarray}
and for terms of the form $\partial_x (f \partial_y \alpha)$,
\begin{eqnarray}
\partial_x (f \partial_y \alpha) = \frac{1}{4 \Delta x \Delta y}
      (f_{i+1,j,k} (\alpha_{i+1,j+1,k}-\alpha_{i+1,j-1,k}) -
       f_{i-1,j,k} (\alpha_{i-1,j+1,k}-\alpha_{i-1,j-1,k})).
\end{eqnarray}
The equation for solving the Hamiltonian constraint for the initial
data would be handled similarly.

Eq. (\ref{fdalp}) forms a set of $N$ equations in $N$ unknowns, where
$N$ is the total number of grid zones, that can be solved for the
values of $\alpha$ at each grid point.  The standard form for a set of
linear equations is that of a matrix equation {\bf Ax}$=${\bf b},
where {\bf A} is an $N \times N$ square matrix containing all the
finite difference coefficients, {\bf x} is a vector of $N$ elements
consisting of all the unknowns and {\bf b} (the right-hand side) is a
vector containing all the source terms of the differential equation.
In this way we generate a standard matrix form of the problem where
{\bf A} is now a very sparse matrix with diagonal structure as shown
in Fig.~\ref{fig:19diag}. Each of the diagonals corresponds to a
finite difference coefficient in the nineteen-point stencil.  For
details on this procedure see Press, et al.~\cite{Press86}.  Other
details of our solver can be found in appendix A or in
Ref.~\cite{Towns93}.

\section{3D Evolution of Black Holes}
\label{sec:evolution}
At present the evolution of 3D black holes is very difficult because
of problems with boundary conditions and resolution requirements. In
this paper we outline these difficulties and present techniques we
developed to overcome them.  The boundary conditions on the black hole
throat are generally provided by an isometry condition that maps the
black hole exterior to a geometrically identical interior sheet, and
these can be troublesome if the coordinate system does not naturally
match the throat. (See Sec.~\ref{sec:isometry} below, where we show
that one can either evolve the entire domain, including the region
interior to the throat, or just the exterior region, when we use an
isometry condition to provide boundary data inside the black hole
throat.)  The outer boundary conditions for black hole evolution are
often taken to be static~\cite{Bernstein93b}, but this is acceptable
only if the outer boundary is sufficiently far away, a condition that
is difficult to obtain in an evenly spaced Cartesian 3D grid with
present computer memories.  The resolution requirements are quite
severe, as has been stressed in previous 1D and 2D studies of black
holes~\cite{Bernstein93b,Seidel92a}.  Whenever a singularity avoiding
lapse is used, large gradients in the metric functions develop due to
the pathological nature of such a slicing condition~\cite{Seidel92a}.
The evolution of the system is frozen inside the horizon, while it
marches ahead just outside, leading to severe stretching of the
coordinates and sharp peaks in the metric.  An example of this effect
is shown in Fig.~\ref{fig:1dgrr}, where the radial metric function is
shown at various times for a maximally sliced black hole (see below
for definitions and more discussion of this point.)  This behavior
often leads to instabilities in 1D and 2D beyond about $t=100M$,
depending on the resolution, and the problem can be more pronounced in
3D because of the limited resolution that can be achieved.

In this paper we compare extensively the results obtained using the 3D
Cartesian code with those obtained using the 1D
codes we previously developed.  Such comparisons are very important in
understanding the effect of various boundary condition and resolution
effects in the 3D code.

\subsection{Black Hole Initial Data}
\label{sec:initial}

The spherical initial data set that we consider in this paper is the
Schwarzschild spacetime represented by a single, Einstein-Rosen
bridge.  This construction is discussed in detail in
Ref.~\cite{Bernstein89}.  The initial 3-metric is given by
\begin{equation}
\label{initdata}
ds^2 = \psi^4 (dr^2 + r^2(d\theta^2 + sin^2\theta d\phi^2)),
\end{equation}
where the conformal factor is $\psi = (1+\frac{M}{2r})$.  Here $r$ is
the isotropic radius, related to the standard Schwarzschild radius
$r_s$ by $r_s = (1+\frac{M}{2r})^2 r$. Transforming to Cartesian
coordinates, we have
\begin{equation}
\label{cartmetric}
ds^2 = \psi^4 (dx^2 + dy^2 + dz^2),
\end{equation}
where the Cartesian coordinates $x$, $y$, and $z$ are related to the
isotropic radius $r$ in the usual way.

In this paper, we shall concentrate on this spherical black
hole initial data set.  The study of other 3D data sets representing a black
hole distorted by a Brill wave~\cite{Bernstein94a} and two colliding
black holes~\cite{Anninos93b,Anninos94a,Anninos94b} will be reported
elsewhere.

\subsection{The Grid}
\label{sec:grid}

The present version of the code is written with a fixed coordinate
grid with equal spacing $\Delta x = \Delta y = \Delta z$ in the
Cartesian coordinate labels.  Because the conformal factor $\psi$ is
singular at the origin, we usually use a grid which straddles the
origin (and coordinate axes), so that the coordinate axes are offset
from grid zones by a half zone, located midway between them.  However,
the code can also place grid zones coincident with the coordinate
axes if desired.  In this case we can set the coordinate values of the
origin to be very small but finite to avoid numerical overflows. Results
presented in this paper have been computed with the staggered grid.

Before going into the details of the calculations, it is instructive
to compare the current Cartesian grid with the logarithmic $\eta$ grid
used in other numerical work on black
holes~\cite{Bernstein93b,Bernstein89}.  In that system, a radial
coordinate $\eta$ is defined by
\begin{equation}
r=\frac{M}{2} e^{\eta},
\end{equation}
where $r$ is the Schwarzschild isotropic radius and $M$ is the
Schwarzschild mass of the black hole.  This coordinate has the
advantage of providing fine resolution near the throat of the black
hole and also near the peak that develops in the radial metric
function, while also allowing the outer boundary to be placed far from
the hole.  Typical high resolution calculations in 1D and 2D based on
this coordinate use $\Delta \eta = 0.03$, with the outer boundary
placed at $\eta = 6$, or $r\approx 200M$.  Low resolution calculations
are performed with $\Delta \eta = 0.06$. Disadvantages of this
coordinate are that: ({\em i}) The throat region remains extremely
well resolved, even after the horizon has moved significantly away
from the hole. Therefore much computational effort is wasted well
inside the horizon where the lapse is typically near zero and the
region is causally disconnected from the outside. ({\em ii}) The grid
becomes very coarse outside the horizon in the radiation zone, because
equal spacing in the $\eta$ coordinate leads to larger and larger
spacing in the more physical $r$ coordinate.  Under these conditions
waves may be reflected back toward the black hole as they are
scattered off of the coarse grid at larger radii, as discussed in
Refs.~\cite{Bernstein93b,Abrahams92a}.

One can estimate what kind of resolutions will be necessary to solve
the spherical black hole problem in 3D by doing runs with a
spherically symmetric (1D) code in $\eta$ coordinates.  It is
instructive to study a high resolution 1D black hole run
($\Delta\eta=0.03$) with maximal slicing to $t=50M$, as shown in
Fig.~\ref{fig:1dgrr}.  At that time, the peak in the radial metric
function is located at approximately $\eta=2$. This corresponds to an
isotropic radius $r$ of approximately $3.7M$. The effective resolution
at the peak in isotropic radial coordinates is thus $\Delta r \approx
0.1M$. Given that this peak is the sharpest feature in the domain, one
might expect that it would be possible to obtain reasonable results
with a 3D code with $\Delta x = \Delta y = \Delta z \approx 0.1M$.  We
also have a 1D code which uses isotropic radial coordinates, which are
more closely related to our 3D Cartesian coordinates.  Experiments
with this code suggest that a resolution of $\Delta r \approx 0.05M$
is actually needed to obtain reasonably accurate results to a time of
$50M$. This difference is due to the fact that although the equally
spaced $\eta$ coordinates cover the throat region extremely well,
where the peak begins to develop at early times, the equally spaced
$r$ coordinates do not, and therefore higher resolution is required.
There are also important geometric factors to consider in 3D, such as
the length of the diagonal of a Cartesian cube being $\sqrt{3} \Delta
x$, which make the resolution requirements more stringent yet.  We
show the size of the black hole throat on a grid of typical resolution
($\Delta x = 0.15M$) in Fig.~\ref{fig:grid}.  In this figure the
throat is located at $r=0.5M$, so we see that the throat itself is not
extremely well resolved. (Recall that $r$ denotes the isotropic, not
Schwarzschild radius.) We show below why it is not necessary to have
the throat highly resolved in many cases.

As mentioned above, the 1D code with $\eta$ coordinates was run with
an $\eta_{max}=6$, which corresponds to $r_{max} \approx 200M$. With
equally spaced Cartesian zones such a luxury is impractical with
present computer memories and speeds: the resolution $\Delta r =
0.05M$ recommended above would require $4000^3$ zones.  Experiments
with 1D codes show that if one is interested in following the metric
and extrinsic curvature components only to a time of $50M$, the outer
boundary should be placed at a radius of greater than about $30M$.
When the outer boundary is too close, its influence will be felt by
the interior solution if the boundary conditions are not properly
formulated, producing error in the height of the peak.  Common
treatments of the outer boundary involve holding the metric functions
fixed or extrapolating them to the outer zones \cite{Bernstein93b},
but these are adequate only when the boundary is placed quite far
away.  When the boundary is placed at $r_{max}=30M$, by $t=50M$ the
error in the peak of $\gamma_{rr}$ (reconstructed from the six
Cartesian metric functions) due to boundary effects (using maximal
slicing) can be about 10\%, depending on the treatment of the outer
boundary conditions. These effects clearly point to the need for
appropriate outer boundary conditions for black hole spacetimes, which
we discuss in section \ref{sec:bc}.

Although experiments with 1D codes allow us to estimate how much
resolution will be needed, they are incomplete guides to the 3D
problem. First of all, resolution issues will be different in
Cartesian coordinates. In 1D and 2D (axisymmetric) codes, constant
coordinate lines are essentially perpendicular to the gradients, so
that symmetries are easily preserved. However, with a Cartesian grid
the constant coordinate lines cross developing features at all angles,
causing finite difference errors to be larger than a 1D treatment
would show. Thus, one expects that even better resolution will be
needed than that implied by the 1D code tests. Also, when considering
how far out to put the boundary of a Cartesian grid, one must take
into consideration that the outer boundary does not have the shape
which is characteristic to the problem (it is a cube, not a
sphere). Thus, boundary conditions must be applied carefully there.
However, in spite of these problems we believe that Cartesian
coordinates are to be preferred over other specialized coordinate
systems.  The symmetries of this first 3D black hole problem are
artificial.  A general 3D spacetime with multiple black holes
will not suggest a preferred coordinate system, as a single hole
does, so a Cartesian mesh should be just as effective as a spherical
or boundary fitted coordinate system (e.g. \v{C}ade\v{z}
coordinates~\cite{Anninos94a}), at least as far as the near zone is
concerned. Furthermore, the singularity free nature of Cartesian
coordinates is clearly desirable.  All axisymmetric numerical
relativity codes (written in either $\rho-z$ or $r-\theta$ type
coordinates) that we are aware of have difficulties near the symmetry
axes and origin (see, e.g., discussion in Ref.~\cite{Bernstein93b}),
and in 3D these problems are much more severe~\cite{Stark91}.

Present memory available on the NCSA 512 node CM-5 (16 GBytes) allows
calculations of up to about $200^3$ zones with our present code (see
the Appendix for discussion of memory requirements for 3D numerical
relativity.)  With the above discussion in mind, it is clear that in
3D black hole simulations, balancing the demands of high resolution
with the need to place the outer boundary sufficiently far from the
hole are quite difficult with a fixed, equally spaced grid.

We have explored a number of techniques to deal with these
difficulties, including ({\em i}) testing better boundary conditions
that allow one to move the boundary closer to the hole, thereby
increasing the affordable resolution, ({\em ii}) using a variable grid
spacing or adaptive mesh refinement (AMR) to add resolution where it
is needed, and ({\em iii}) using an apparent horizon boundary
condition to remove the peak from consideration, thereby reducing the
resolution requirements dramatically.  The boundary conditions and
apparent horizon shift are discussed in sections \ref{sec:bc} and
\ref{sec:shift} respectively.  The use of variable meshes and dynamic
adaptive mesh refinement will be reported elsewhere.

\subsection{Boundary Conditions}
\label{sec:bc}

In this section we discuss our choice and implementation of boundary
conditions.  Because we are evolving a spherical black hole, it is not
necessary to evolve the entire system.  In order to achieve the
highest possible resolution, we often choose to place the black hole
at the origin of our coordinate system and evolve only a single octant
of it.  Then the boundary conditions on the $x=0$, $y=0$, and $z=0$
planes are given by the symmetry of the spacetime, as discussed below.
Note that this treatment can be extended to all spacetimes which have
both axial and equatorial symmetry, so that all of the NCSA
axisymmetric black hole studies performed to date, including the
collision of two equal mass black holes, can be studied with this
geometry.  As we report below, we also have performed simulations
where the entire domain is evolved, and comparisons with the
evolutions performed with symmetry conditions show identical results,
as expected.

\subsubsection{Black hole isometry}
\label{sec:isometry}

Application of boundary conditions on the inner surface that is the
black hole throat is made difficult by the choice of an ``unnatural''
grid. Cartesian grids do not conform to the spherical black hole
surface and as a result the formulation of accurate finite differenced
conditions at the throat is considerably more involved than would be
the case if we had adopted a spherical coordinate system as we have in
previous 2D work \cite{Anninos93c,Anninos93b,Bernstein93b}.

We use an Einstein-Rosen bridge construction to connect two
asymptotically flat sheets and form a black hole. This
construction provides boundary conditions by allowing the use of an
isometry to map the metric exterior to the throat (or isometry surface)
to the interior regions. The isometry conditions take the form
of a map $J$ which identifies the two sheets through the throat
\cite{Cook90}:
\begin{equation}
\gamma_{ab}(\vec{x})=\pm J^c_a J^d_b \gamma_{cd}(J(\vec{x}))
\label{isom1}
\end{equation}
with
\begin{equation}
J(\vec{x})=a^2\frac{\vec{x}-\vec{c}}{\left|\vec{x}-\vec{c}\right|^2} +
\vec{c}
\end{equation}
and $J_a^b =\frac{\partial J^b}{\partial x^a}$,
where $a$ is the radius of the black hole throat centered at $\vec{c}$.
The mapping (\ref{isom1}) is applicable to both the metric and
extrinsic curvature tensor fields.

This isometry technique has been used extensively in the construction
of black hole initial data sets~\cite{Cook93,Bowen80,Bernstein94a},
but can also be used to provide boundary conditions during the
evolution of black hole spacetimes.  For example, in all previous NCSA
2D evolutions of distorted~\cite{Abrahams92a},
rotating~\cite{Brandt94a,Brandt94b}, and
colliding~\cite{Anninos93b,Anninos94a,Anninos94b} black holes, a
coordinate system was chosen so the isometry condition on the 3-metric
took on a very simple form: $\eta \rightarrow -\eta$, where $\eta$ is
a radial coordinate.  The evolution equations themselves also respect
this symmetry if the lapse, shift, and extrinsic curvature variables
obey the isometry as well.  Generally, the radial component of the
shift must vanish on the throat, and the lapse and extrinsic curvature
components must have the same isometry sign (positive or negative).
Under such conditions all evolution and constraint equations are
preserved across the throat.  The same situation occurs in the 3D
case, as long as one is careful to apply proper boundary conditions on
all variables.  As we discuss in sections \ref{isom} and
\ref{sec:results} below, it is also possible to {\em evolve} the data
inside the throat, so that the isometry condition need not be applied
if it is not desired.

The change of sign in Eq. (\ref{isom1})
results from the square of the map being an identity
so that $J$ is its own inverse and is chosen from the
continuity and consistency constraints of the Einstein equations.
For example, the metric must obey the isometry with a plus sign
to be nonsingular on the throat and we define the lapse condition as
\begin{equation}
\alpha(\vec{x})= \pm \alpha(J(\vec{x}))
\label{isom2}
\end{equation}
with the sign taken to be the same as that of the extrinsic curvature.
In the work presented in this paper we have used time slices in which
the lapse is either symmetric or antisymmetric across the throat so
that either of the isometry signs in (\ref{isom2}) may be used.

If the metric is conformally flat so that $\gamma_{ab}=\psi^4 f_{ab}$
where $f_{ab}$ is the flat space background, we get from (\ref{isom1})
\begin{equation}
\psi(\vec{x})=\frac{a}{r}\psi(J(\vec{x}))
\end{equation}
for the conformal factor. Because our initial data is defined to be
conformally flat and we fix $\psi$ to be constant in time,
we compute the conformal factor at the first time slice and
apply the isometry to the conformal metric in subsequent time slices.
Denoting $\gamma_{ab} =\psi^4 \bar{\gamma}_{ab}$, we can write the
isometry on $\bar{\gamma}_{ab}$ as
\begin{equation}
\bar{\gamma}_{ab}(\vec{x})=\left(\frac{r}{a}\right)^4 J_a^c J_b^d
                     \bar{\gamma}_{cd}(J(\vec{x})),
\label{isomconf}
\end{equation}
where $\bar{\gamma}_{ab}$ is the conformal metric. We found this
construction to be more accurate and stable than performing the
isometry to the full unconformed metric components, since the
conformal factor is known analytically and the ``flat'' metric
functions can be more accurately interpolated. Note that this
important trick can be used even if the conformal factor is known only
numerically.

Boundary conditions on $\bar{\gamma}_{ab}$ at the throat can be
derived by differentiating equation (\ref{isomconf}) and taking the
limit $r\rightarrow a$ (see for example reference \cite{Cook90}). An
alternative construction that we have developed is to compute the
isometric coordinates $J(\vec{x})$ for $\left|\vec{x}\right|<a$ and
evaluate the corresponding tensor components inside the throat by
volume weighting the nearest isometric neighbors as
\begin{equation}
\bar\gamma_{ab}(\vec{x})=\left(\frac{r}{a}\right)^4 J_a^c J_b^d
     \sum_{n=1}^{8} \bar\gamma_{cd}
\left(J^{(n)}(\vec{x})\right)\Delta V^{(n)},
\end{equation}
where the index $n$ refers to the eight cells overlapping a cube of size
$(\Delta x,\ \Delta y,\ \Delta z)$ centered around $J(x)$. $V^{(n)}$ are
the corresponding volume weights.
In this way we solve algebraic identities
and not differential relations across the throat which would require a complex
network of discrete stencils and logical switches as discussed
in reference \cite{Cook93}.

\subsubsection{Other boundaries}
\label{sec:otherbc}

Uniform grid spacing and limitations on available computer time and
memory severely restrict the placement of the outer boundaries.
Ideally we would like to impose asymptotic and conformal flatness
$\psi, \bar{\gamma}_{ab} \rightarrow 1$ as $r \rightarrow \infty$. One
would expect that in 3D, where the outer boundary is not far enough
away from the throat, neither asymptotic flatness nor static outer
boundaries would be sufficient. However, we found that for most of our
runs, static outer boundaries worked better than any extrapolation
boundary condition.

For the runs which used an apparent horizon locking shift, however,
one extrapolation outer boundary condition worked well.  The method
involves matching a Schwarzschild-like solution to the outer boundary
zones and is applicable to the diagonal metric components. This is
done by defining an effective mass ``constant'' $\tilde{K}$
\begin{equation}
\tilde {K}_{io-1}=r_{io-1}\left(\gamma_{io-1}^{1/4} - 1\right)
\end{equation}
independently for the outermost zones (labeled by the index $io-1$)
that still live inside the computational domain. We then construct the
boundary condition by extending the Schwarzschild-like solution to the
outside boundary zone as
\begin{equation}
\gamma_{io} = \left(1 + \frac{\tilde{K}_{io-1}}{r_{io}}\right)^4.
\end{equation}

Another kind of boundary in our code comes from our evolving only one
octant of the system.  Boundary conditions must also be supplied for
the planes $x=0$, $y=0$, and $z=0$. The setting of boundary condition
there is straightforward as they are determined by the symmetries of
the problem
\begin{eqnarray}
0&=&\gamma_{xy}|_{x=0}
  = \gamma_{xy}|_{y=0}
  = \gamma_{xz}|_{z=0} \nonumber \\
 &=&\gamma_{xz}|_{x=0}
  = \gamma_{yz}|_{y=0}
  = \gamma_{yz}|_{z=0}
\end{eqnarray}
and
\begin{eqnarray}
0&=&\partial_x\alpha |_{x=0}
  = \partial_y\alpha |_{y=0}
  = \partial_z\alpha |_{z=0} \nonumber \\
 &=&\partial_x\gamma_{xx}|_{x=0}
  = \partial_y\gamma_{xx}|_{y=0}
  = \partial_z\gamma_{xx}|_{z=0} \nonumber \\
 &=&\partial_x\gamma_{yy}|_{x=0}
  = \partial_y\gamma_{yy}|_{y=0}
  = \partial_z\gamma_{yy}|_{z=0} \nonumber \\
 &=&\partial_x\gamma_{zz}|_{x=0}
  = \partial_y\gamma_{zz}|_{y=0}
  = \partial_z\gamma_{zz}|_{z=0} \nonumber \\
 &=&\partial_x\gamma_{yz}|_{x=0}
  = \partial_y\gamma_{xz}|_{y=0}
  = \partial_z\gamma_{xy}|_{z=0}
\end{eqnarray}
The extrinsic curvature
components obey identical conditions as the corresponding metric
components.

We stress that our code is not restricted to the use of these boundary
conditions, and that the black hole may be placed in the center of the
grid if desired.  In section \ref{sec:results} below we present
examples of such calculations which show that the results are
identical whether the black hole is placed in a corner of the grid or
in the center, as they should be.  We are exploiting this symmetry in
order to achieve the highest possible resolution, while still treating
the black hole as a true 3D system in one octant. This treatment
allows us to study 3D evolution with effectively eight times the
memory than we would have with a black hole at the center of the grid.

\subsection{Computational Considerations}
\label{sec:tricks}

In this section we discuss a few important computational issues
regarding the evolution of 3D black holes in Cartesian coordinates.
We have developed various techniques to make black hole evolution
calculations accurate and stable.  Specific examples of results
obtained using these techniques are provided in the sections that
follow.

\subsubsection{Conformal Derivatives}
\label{conformal}
The conformal factor $\psi$ is usually obtained by solving the
Hamiltonian constraint on the initial time slice in numerical
relativity calculations.  For black hole initial data, this function
often peaks up near the throat of the black hole(s).  For our initial
data set, $\psi$ is known analytically to be $\psi = 1 +
\frac{M}{2r}$.  Since the metric functions are related to $\psi^4$,
they can become quite steep near the throat.  In 3D Cartesian
coordinates, it can be difficult to afford high resolution in this
region of the spacetime if one also wishes to cover the radiation zone
away from the hole, so derivatives of the full metric functions can be
particularly inaccurate if special care is not taken to compute them.

We have dealt with this problem by constructing the conformal factor
$\psi$ and all its spatial derivatives on the initial time slice and
storing them in memory with machine accuracy.  Then, while the full
metric functions (e.g. $\gamma_{xx}$) are evolved, their spatial
derivatives are computed by taking the conformal factor explicitly
into account.  The conformal metric functions are computed by dividing
out the conformal factor. Derivatives of these (more slowly varying)
conformal functions are computed numerically, and the full metric
derivatives are constructed by substituting the appropriate stored
``analytic'' values for the conformal factor and its derivatives.
This procedure, which we call ``conformal differentiation'', proves
critical in obtaining accurate and stable evolutions particularly near
the throat. It should be noted that this technique will be useful even
when we have numerically generated initial data, because the initial
data problem for the conformal factor $\psi$ can be solved with very
high accuracy, and then its numerically computed derivatives can also
be known as accurately as desired on the grid used for the evolution.
Experimentation shows that it is most important to apply this
technique to the diagonal 3-metric functions.

\subsubsection{Isometry}
\label{isom}
In section \ref{sec:isometry} above we discussed the implementation of
the isometry condition on the black hole throat.  This condition
provides a good boundary condition on the inner region of the black
hole spacetime.  It forces the spacetime interior to the throat, in
the other ``universe'', to have the same geometry as that outside the
the throat, where the spacetime is actually evolved, and its
application allows us to evolve only the region on and outside the
black hole throat.

However, this is simply a choice that one can make as a matter of
convenience in evolving the spacetime.  Alternatively one could evolve
both sides of the throat as one sees fit.  In our 3D Cartesian
coordinate system, this is an easy option.  The initial data are known
everywhere, both inside the throat and out, so in principle one can
simply evolve the entire spacetime without appealing to an isometry
condition on the black hole throat.  Although the region inside the
throat is very poorly resolved and the spacetime will not be very
accurate there, this need not pose a problem since we are really only
interested in the exterior region.

As we discuss in the sections below on the evolution of the black
hole, we decide whether or not to apply the isometry depending on the
behavior of the chosen lapse in the region of interest.  If one is
interested in evolving the black hole with geodesic slicing, where
large gradients will develop near the throat, it is important to use
the isometry boundary condition for accurate results that can be
compared with 1D codes.  For some singularity avoiding lapses that
tend to collapse quickly, the isometry need not be imposed and the
entire spacetime, inside and outside the throat, can be evolved
without problems (until late times when large peaks develop in metric
functions near the horizon).  In these cases the lapse tends to
collapse near the black hole throat and also inside it, so that the
evolution is essentially frozen in this region.  For lapses that have
the negative isometry sign initially, we have found it to be important
to use the isometry throughout the evolution because of large
gradients that develop in the lapse near the origin, as we discuss in
more detail in section \ref{sec:results} below.  Finally, we point out
that we are ultimately interested in applying an apparent horizon
boundary condition in 3D black hole spacetimes, which in principle
obviates the need to consider any treatment of the
throat~\cite{Anninos94e}.  In section \ref{sec:ahresults} below we
will show an example of how this can be achieved in 3D.

\subsubsection{Numerical Viscosity}
\label{viscosity}
Another important problem that occurs when evolving black hole
spacetimes is related to effects of singularity avoiding time slicing
conditions. Large peaks develop in the vicinity of the black hole
horizon as time slices push forward away from the hole but are held
back inside it (see, e.g., Fig.~\ref{fig:1dgrr}).  Sharp peaks that
develop in the solution to hyperbolic equations can cause numerical
instabilities~\cite{Press86}, as we have seen in the black hole
problem.  A common strategy to cope with this problem is to add a
small diffusion term to the evolution equations that effectively
smooths out short wavelength features (such as numerical noise).  The
usual way of achieving this is to add a second derivative term to an
evolution equation with a small coefficient in front.  In this way,
for example, the evolution for the metric functions would be
\begin{equation}
\partial_{t}\gamma_{ab}=-2\alpha K_{ab}+
D_{a}\beta_{b}+D_{b}\beta_{a} + \epsilon \nabla^2 \gamma_{ab},
\end{equation}
where $\epsilon$ is very small.  As we discuss below, this
technique is important for the lapse evolution used for algebraic
slicings but it has not cured the difficulties associated with grid
stretching.  It does damp out some noise at early times, but at late
times when peaks are poorly resolved, it is not adequate.

\subsection{Lapse}
\label{sec:lapse}
It is well known that the Einstein equations do not determine either
the lapse function or the shift vector. These quantities may be chosen
freely.  As nearly all work to date in numerical relativity has been
done in 1D or 2D, most lapse and shift conditions have been developed
with a particular symmetry or coordinate system in mind.  In 3D
Cartesian coordinates, with no symmetries, these gauge and slicing
conditions must be reexamined.  In this section we discuss standard
slicing conditions and their use in 3D, and propose another class of
algebraic slicings that seem especially suited to 3D work.

When evolving black holes numerically, the choice of lapse function is
motivated by the need to keep the numerical grid from falling into the
singularity. It has been shown that maximal slicing has this
singularity avoiding property
\cite{Estabrook73,Lichnerowicz44,Reinhardt73,Eardley79,Smarr78a}, so
this slicing condition has been the most frequently used in 1D and 2D
calculations. However, using maximal slicing requires solving an
elliptic equation, which is computationally expensive, especially in
3D. Therefore, as we report below, we have also used a number of
algebraic slicings which mimic maximal slicing.

\subsubsection{Geodesic Slicing}
\label{sec:geodesic}
A very strong test of the code can be made by evolving the black hole
with geodesic slicing, or simply $\alpha = 1$ and $\beta^a=0$.  With
this slicing condition one can show (see, e.g., \cite{Misner73}) that
a point initially on the black hole throat must fall into the
singularity after a proper time (in this case identical to coordinate
time) $\tau=t=\pi M$.  In a numerical evolution, this is manifested by
the radial metric function $\gamma_{rr}$ approaching $\infty$ and the
angular metric function $\gamma_{\theta\theta}$ going to zero as a
true curvature singularity develops.  In short, the code must crash at
$t=\pi M$.  But particularly as all the various $Cartesian$ metric
functions are evolved, not the just spherical metric functions
$\gamma_{rr}$ and $\gamma_{\theta\theta}$, and the isometry routine is
used to provide boundary conditions on the throat as grid points crash
towards the singularity, this is a serious code test.  The system must
remain spherical, even as singular structures are developing in the
various Cartesian metric functions.

Geodesic slicing also allows us to compare data from the 3D code with
that obtained from a spherically symmetric 1D code without the
complication of a lapse computation. See section \ref{sec:geodresults}
for both the crash time tests and the 1D comparisons.

\subsubsection{Maximal Slicing}
\label{sec:maximal}

Maximal slicing has been used extensively in numerical relativity
(see, e.g.,
\cite{Estabrook73,Evans86,Bernstein89,Anninos93c,Anninos94b}) for
several reasons: considerable analytic work has been done delineating
its excellent singularity avoiding properties
\cite{Estabrook73,Lichnerowicz44,Reinhardt73,Eardley79,Smarr78a}, it
was used in early numerical work which centered on computing black
hole spacetimes with Einstein-Rosen bridges, and it is conveniently
computed from the 3--metric and extrinsic curvature.  Maximal slices
(i.e. hypersurfaces with maximal volume) are characterized by the
vanishing of their mean curvature
\begin{equation}
\label{trkzero}
\mbox{tr} K \equiv \gamma^{ab} K_{ab}=0.
\end{equation}
Inserting this condition on the evolution equation for ${\mathrm tr} K$
yields a condition on the lapse
\begin{equation}
\label{maximal slicing1}
D^{a}D_{a} \alpha = \alpha R,
\end{equation}
or, using the Hamiltonian constraint,
\begin{equation}
\label{maximal slicing2}
D^{a}D_{a} \alpha = \alpha K^{ab} K_{ab}.
\end{equation}
This latter form, fully expanded in Eq. (\ref{maximal slicing}), is
the form used in the code.  The right-hand-side of Eq.~(\ref{maximal
slicing1}) contains second derivatives of the metric functions, which
can become difficult to compute accurately when peaks develop at late
times.  The form given by Eq.~(\ref{maximal slicing2}) eliminates
these derivatives and tends to be better behaved numerically.  As
noted in section \ref{sec:elliptic}, we have developed a routine for
the CM-5 which solves this kind of 3D elliptic
equation. The results we obtained using this routine are discussed in
section \ref{sec:maxresults}.

\subsubsection{Algebraic Slicings}
\label{sec:algebraic}
Maximal slicing is an effective time slicing condition for avoiding
singularities, but it suffers from several problems.  It is very
costly, especially in 3D on parallel machines, and it must be solved
to a very high tolerance if noise in the solution is to be avoided, as
we discuss in appendix \ref{sec:appendix} below.  However, slicing
conditions are completely arbitrary, and there are many singularity
avoiding lapse conditions that do not involve solving elliptic
equations.

The most convenient of these slicing conditions for numerical purposes
belong to a class of {\em algebraic} lapse conditions.  These lapse
functions are algebraic combinations of the 3-metric components;
typically they are functions of the determinant of the 3-metric.  One
of the most well known of these slicing conditions is the harmonic
time slicing condition.  In the absence of a shift vector, this
condition reduces to the simple equation
\begin{equation}
\label{harmlapse}
\alpha = f(x^a) \sqrt{\gamma},
\end{equation}
where $\gamma$ is the determinant of the conformal 3-metric and
$f(x^a)$ is a function of the three spatial coordinates to be
specified.  This lapse choice is known to be singularity avoiding, but
just barely so.  It has been shown by Bona and Mass\'o~\cite{Bona88}
that this condition does not allow a time slice to hit a curvature
singularity within a finite coordinate time, but it will come
arbitrarily close.  A separate code that uses this slicing condition
exclusively, based on the formulation of Bona and
Mass\'o~\cite{Bona92}, has been developed~\cite{Bona92b,Anninos94d}.
However, it is not considered here as the weak singularity avoiding
nature of harmonic slicing makes it difficult to apply to black holes
without some sort of apparent horizon boundary condition. Their
formulation has recently been extended to cover all singularity
avoiding slicing conditions considered in this paper, including
maximal \cite{Bona94b}, and a 3D code based on this formulation is
under construction.

Motivated by the harmonic slicing, we have explored a number of
algebraic conditions that are simple to compute and avoid
singularities more strongly than the harmonic condition.  Many such
conditions have been explored by Bernstein~\cite{Bernstein93a} in his
studies of spherical black holes.  Generalizing the harmonic condition
we can choose
\begin{equation}
\label{alglapse}
\alpha = f(x^a) g(\gamma),
\end{equation}
where again $f(x^a)$ is an arbitrary function of the spatial
coordinates and $g(\gamma)$ is some function of the conformal
determinant of the 3-metric.  These functions can be chosen as
desired.  One choice for $g(\gamma)$ that works particularly well
leads to the slicing condition
\begin{equation}
\label{alglapse1}
\alpha = f(x^a) (1 + \text{log}(\gamma)).
\end{equation}
This lapse condition has the remarkable property that it mimics the
action of maximal slicing.  For spherical black holes, even at late
times this slicing condition leads to spatial metric functions
(e.g. $\gamma_{rr}$ and $\gamma_{\theta\theta}$) that are quite
similar in profile and size to those obtained with maximal slicing. A
comparison of the metric radial function obtained with these two
slicings in 1D is shown in Fig. \ref{fig:a1dmaxalg}.  Furthermore, one
can show that if $f$ itself obeys an isometry condition, this lapse
condition has the nice property that it transforms properly under the
isometry operation discussed in section \ref{sec:isometry}, so that
the 3-metric $\gamma_{ab}$, the extrinsic curvature $K_{ab}$, and the
lapse $\alpha$ all transform across the throat in such a way as to
preserve the evolution equations.

The algebraic slicings do have one notable drawback relative to
maximal slicing in that they tend to be a bit too ``local''.  If some
feature develops at a particular point in the 3-metric, the algebraic
lapse responds instantly and locally.  This is then fed back into the
evolution equations through first and second spatial derivatives of
the lapse, which can exaggerate undesirable features in the solution.
The solution to an elliptic equation, on the other hand, tends to
smooth over any local inhomogeneities.  In order to decrease the
locality of the algebraic lapse conditions, we have taken advantage of
the evolution equation for the determinant of the 3-metric and used it
to actually evolve the lapse.  Then we may add a diffusion term to the
evolution equation for the lapse.  The diffusion tends to smooth out
any local, higher frequency features that develop, leading to a more
stable evolution.  This approach leads to the following scheme for the
lapse:
\begin{equation}
\label{evolvelapse}
\dot{\alpha} = f(x^a) g'(\gamma) \dot{\gamma} + \epsilon \nabla^2 \alpha,
\end{equation}
where $\dot{\gamma}$ is given by
\begin{equation}
\label{evolvedet}
\dot{\gamma} = \gamma (-2 \alpha trK + 2 D_a \beta^a).
\end{equation}
With lapse choices like this we were able to evolve the spherical
black hole to nearly $50M$ in time, as discussed in section
\ref{sec:algeresults}.

\subsection{Shift Vector}
\label{sec:shift}

For the most part the simulations presented in this paper are
performed using a vanishing shift vector. However, the presence of
black holes in numerical spacetimes introduces extreme behavior in the
metric variables leading to large gradients that inevitably develop in
the vicinity of the horizon when traditional singularity avoiding
lapses are used.  In 1D and 2D codes these problems have been
troublesome, but not insurmountable, as the coordinate systems have
generally been well suited to the problem at hand.  In 2D, the trouble
is essentially reduced to a 1D problem since the gradients occur
primarily in the radial direction.  For this reason, in these systems
one could afford to add enough radial zones to solve or tame the
problem by brute force for a certain period of time.  However, even in
these cases significant errors (particularly evident in the horizon
mass) emerge from $t\sim 50M$ and grow to severe code crashing
proportions by $t\sim 100M$ \cite{Bernstein89,Anninos94e}.

An approach that is often tried is to use a shift vector to help reduce the
gradients that develop in black hole simulations.  A careful study of
many commonly used shift conditions applied to the Schwarzschild
spacetime has been carried out by Bernstein~\cite{Bernstein93a}.  He
has found that, at least for maximal time slicing, shift vectors such
as minimal distortion, quasi-isotropic (for this spherical case the
quasi-isotropic gauge is the minimal distortion gauge
\cite{Bernstein93a}), minimal strain, and various others, fail for the
spherical black hole.  However, in that study the radial component of
the shift was forced to vanish at the throat of the black hole to
satisfy the boundary condition that the isometry surface remains
there.

In a new approach to evolving black hole spacetimes, the inner
boundary of the computation is chosen not to be the throat of the
wormhole, but the apparent horizon~\cite{Thornburg93} or some point
just inside the apparent horizon~\cite{Seidel92a,Anninos94e,Bona94a}.
Data inside this region, including the singularity, are causally
disconnected from the region outside the horizon and are simply
deleted from the calculation.  For this reason, a singularity avoiding
lapse is not required for black hole evolution; the singularity is
avoided by removing it from the problem.  Such an apparent horizon
boundary condition has shown {\em dramatic} improvements in evolving
black hole spacetimes in 1D, allowing evolutions of order $t\sim
1000M$ with errors of order a few percent or
less~\cite{Seidel92a,Anninos94e,Bona94a}.  We believe this sort of
apparent horizon shift condition will be essential to the development
of accurate and stable 3D black hole codes in the future.

In Ref.~\cite{Anninos94e} a number of shift vectors were introduced
and studied for the realization of the apparent horizon boundary
condition.  Although the location of the horizon is independent of the
shift vector on any particular time slice, the time rate of change of
its coordinate position is not.  Hence we can determine the value of
the shift at the horizon that is needed to keep the horizon from
moving across coordinates by, e.g.,
\begin{equation}
\left.\partial_t \Theta (\vec{x})\right|_{\vec{x} = \vec{x}_{AH}}=0~,
\end{equation}
where $\vec{x}_{AH}$ is the horizon position and
\begin{equation}
\Theta = D_a s^a + K_{ab} s^a s^b - K
\end{equation}
is the expansion of the outgoing null rays on the spacelike 2-surface
that defines the apparent horizon ($\Theta = 0$) with unit outward
pointing 3-vector $s^a$ \cite{York89}.  In addition to defining a
local shift vector to control the motion of the horizon, we must also
specify values for the shift at all other points in the spacetime.
Several such constructions were presented in Ref.~\cite{Anninos94e},
including a ``distance freezing'' shift, an ``area freezing'' shift,
and the ``minimal distortion'' shift.  Here we adopt one of these
constructions, namely the ``distance freezing'' shift, to explore the
feasibility of implementing horizon locking coordinates in 3D. In
particular, we require that the radial metric function remain constant
in time such that $\partial_t \gamma_{rr}=0$.  In spherical geometry
this yields a first order differential equation for the radial shift
vector that can be solved as a single boundary value problem with the
boundary condition set to lock the horizon in place.  Of course as the
problem is treated as a full 3D problem, we use the three Cartesian
shift components $\beta^x$, $\beta^y$, and $\beta^z$ in this
construction.  Results using this particular shift vector are
discussed in more detail in Sec.~\ref{sec:ahresults}

\section{Results}
\label{sec:results}

\subsection{Geodesic Slicing}
\label{sec:geodresults}

As mentioned above in section~\ref{sec:geodesic}, when a Schwarzschild
black hole is evolved with geodesic slicing, the analytic solution
requires that the time slice hit the singularity at a proper time of
$t = \tau = \pi M$. Many runs were made at different resolutions to
test our code against this analytic result. In
Fig.~\ref{fig:crashtimes} we show the time required to hit the
singularity for a set of runs in which all parameters are held
constant except for the spatial resolution and the Courant factor.
Clearly, as either the grid spacing or the Courant factor is
decreased, the ``crash time'' approaches $\pi M$.  There are two
effects at work here.  First, as one increases the resolution, the
code is able to more accurately resolve the throat region where metric
functions are becoming singular.  Second, as the time step is made
smaller one is able to resolve the time at which the slice hits the
singularity with more precision.

Evolving the black hole with geodesic slicing also allows one to
perform a serious test of the evolution equations without the
complication of the lapse computation. When one runs the code with
high spatial resolution, e.g. $\Delta x=\Delta y=\Delta z=0.025 M$,
the peaks in $\gamma_{rr}$ and $\gamma_{\theta\theta}$ obtained from
the Cartesian metric functions line up almost exactly with those
produced by a spherical, 1D code run with the same resolution, as
shown in Figs.  \ref{fig:geodcompa} and \ref{fig:geodcompb}.  These
functions have been constructed from the Cartesian metric function
that are actually evolved.  In Fig.~\ref{fig:geogxx} we show a 2D
slice through the plane $z=0$ of the Cartesian metric function
$g_{xx}=\gamma_{xx}/\psi^4$ evolved to a time $t=3M$.  (We factor out
the dependence of the conformal factor $\psi$ to show the dynamical
evolution more clearly.)  Along the $x$-direction, this function
behaves like $g_{rr}=\gamma_{rr}/\psi^4$, but in other directions it
does not.  It is only through the combination of all Cartesian metric
functions that the spherical behavior of $g_{rr}$ is seen.

As mentioned above, it is not necessary in principle to apply
the isometry condition across the throat. We have shown, however, that
for accurate evolution it is necessary for the geodesic slicing case.
When the isometry is not applied, the peak in the radial metric
function grows faster than it should, causing the code to crash
earlier than it would otherwise.  Also, the agreement with the 1D code
is much better when the isometry condition is applied.  This should
not be surprising, because the isometry condition maps the entire
spacetime region exterior to the throat to the domain inside the
throat.  As we are covering this interior region with a very small
number of zones, it is impossible to evolve the interior accurately in
this manner.  By mapping the well resolved exterior numerical
solution to the poorly resolved interior, we are able to achieve very
high effective resolution inside the throat.

The 3D results discussed thus far were obtained with the black hole
throat located in a corner of the 3D grid, thereby permitting
simulations that cover only one-eighth of the total spacetime volume.
Symmetry conditions were applied at the faces of the cube that match
up to other interior regions of the spacetime, as discussed in
section~\ref{sec:bc}.  To show that this technique does not affect our
results, we now show results obtained by evolving the full spacetime
domain with the black hole throat located at the center of the 3D
grid.  In Fig.~\ref{fig:geofullgrr}, we show a surface plot of a slice
of the full 3D grid (at $z=0$) of the metric function $g_{rr}$
reconstructed from the Cartesian metric functions.

We also took advantage of the simplicity of geodesic slicing to
perform convergence tests on the evolution equations. Convergence
tests done on the code with second order spatial derivatives show that
our code is indeed second order. A number of metric functions have
been tested with similar results.  Convergence tests of this code
applied to gravitational wave
data sets give similar results~\cite{Anninos94d}.

In addition to comparing directly the behavior of metric functions
obtained with our 3D Cartesian and 1D spherical codes, we have also
compared derived quantities that are sensitive measures of the physics
being computed.  For example, the location and mass of the apparent
horizon of a black hole evolved numerically are particularly sensitive
to errors in the calculations~\cite{Seidel92a,Anninos94b,Brandt94c}.
Although we have not yet connected a full 3D apparent horizon solver
to our code, we can take advantage of the fact that we expect our data
to be spherical. (A 3D apparent horizon finder is under
development~\cite{Libson94b}.)  Therefore we can transform the
Cartesian metric and extrinsic curvature components into their
spherical coordinate counterparts, interpolate them onto a radial line
which intersects the origin, and locate the apparent horizon along
that line using a 1D apparent horizon finder.  Using this method, we
were able to locate the position of the apparent horizon and compute
its mass $M_{AH}$ using the area relation
\begin{equation}
M_{AH} = \sqrt{\frac{A}{16\pi}},
\label{ahmass}
\end{equation}
where $A$ is the surface area of the horizon.  For a Schwarzschild
spacetime, $M_{AH}$ should be equal to the ADM mass of the spacetime
throughout the evolution, but due to numerical error this is difficult
to achieve at late times, even in 1D and 2D
codes~\cite{Seidel92a,Anninos94b,Brandt94c}.  With our 3D code we have
computed these quantities along four different lines (the $x-$, $y-$,
and $z-$ axes and the diagonal).  Using the location of the horizon
found along these lines, we can compute an effective mass at each
point by taking the metric functions found there and computing the
area, assuming spherical symmetry.  Both the mass $M_{AH}$ and the
location of the horizon agree well both along different lines in the
3D domain, and with the values obtained with a 1D code. The position
of the apparent horizon is shown in Fig.~\ref{fig:geodah}.  All lines
are shown, but they are indistinguishable on this plot. In
Fig.~\ref{fig:geodahmass} we plot the apparent horizon mass computed
along all four lines in 3D, as well as the 1D result.  The masses are
all within 0.07\% of each other and the ADM mass at $t=3M$.

\subsection{Maximal Slicing}
\label{sec:maxresults}

We now turn to results obtained using maximal slicing.  These
simulations were all performed by evolving the spacetime both outside
and inside the throat, without using the isometry condition.  Earlier
work on numerical black hole evolution with maximal slicing in 1D
(see, e.g.~\cite{Bernstein89}) and 2D (see,
e.g.,~\cite{Anninos93c,Brandt94c}) has taken advantage of the isometry
to use the throat as an inner boundary on both the evolution and the
solution to the maximal slicing equation.  In those cases, the
isometry was a simple differential condition across the throat in the
spherical coordinate system used.  In this 3D case the Cartesian
boundary conditions on the lapse are cumbersome to apply in the
elliptic solver, so we have chosen to evolve the entire domain inside
and outside the throat.  As we will show below, the lapse collapses
quickly both outside {\em and} inside the throat, halting the
evolution there.  Although in this case the evolution inside the
throat is no longer isometric to that outside, it is of little
consequence.

The first case we consider is a simulation evolving a single octant of
the full 3D spacetime.  This is a typical run, with a resolution of
$130^3$ equally spaced zones, with $\Delta x = 0.1M$.  The outer
boundary was located at $x=12.95M$. Such a calculation can be
performed on a daily basis on the NCSA CM-5.  In this case the
boundary conditions on the metric and extrinsic curvature components
are treated as in section~\ref{sec:otherbc}.  The boundary conditions
on the lapse are treated in a similar manner, with reflection symmetry
used at the inner boundaries ($x=0$, $y=0$, and $z=0$ planes) and the
spherical Schwarzschild value is maintained at the outer edges of the
grid (i.e.  $\alpha = (2r-M)/(2r+M)$).  This treatment at the outer
edges of the cube was crucial in maintaining a stable evolution there.
If the lapse is taken to be, say
$\alpha=1$ in the outer region, serious edge effects and nonspherical
behavior develop quickly.

First we show the lapse function $\alpha$ at a late time of $t=28M$.
In Fig.~\ref{fig:maxalpha} a 2D slice of the lapse through the plane
$z=0$ is shown.  It has collapsed throughout the throat region and
also in a region outside it, and then climbs steeply towards its outer
Schwarzschild value in spherical step function fashion.  A full 3D
analysis indicates that the lapse is quite spherical throughout the
volume, in spite of the fact that our solution is carried out in
Cartesian coordinates with boundary conditions imposed on the faces of
a Cartesian cube.

In Fig. \ref{fig:maxgxx} we show a 2D slice (at the plane $z=0$) of
the conformal metric function $g_{xx} = \gamma_{xx}/\psi^4$ at the
same time $t=28M$.  By this time serious gradients and shearing are
developing in the metric functions due to the grid stretching effects
that result from maximal slicing.  Along the $x-$direction, the metric
function $g_{xx}$ behaves much like a radial metric function $g_{rr}$.
Along the $y-$direction, however, the function $g_{xx}$ is essentially
flat, and along the diagonal there is a very sharp transition region.
This is quite typical of the effects of singularity avoiding lapse
conditions.  In Fig. \ref{fig:maxgrr} we show a 2D slice (through the
plane $x=0$) of the metric function $g_{rr} =\gamma_{rr}/\psi^4$
reconstructed from the six Cartesian metric functions at time $t=28M$
for the same simulation.  The familiar spherical peak is developing
around the black hole, as in 1D and 2D calculations (see, e.g.,
Fig.~\ref{fig:1dgrr}).

As in the previous section on geodesic slicing, we have tracked and
analyzed the apparent horizon in these simulations, and compared the
results to those obtained using a spherical, 1D code.  In
Fig.~\ref{fig:maxah} we show the apparent horizon location from our 3D
code as determined by considering radial lines along the $x-$, $y-$,
and $z-$ axes and the diagonal of the cube.  The independent results
obtained by evolving a maximally sliced spherical black hole with a 1D
code are shown as a dot-dashed line.  The agreement is quite good,
within one 3D grid zone at late times.  We do not expect perfect
agreement in the location of the horizon in coordinate space, since
the slicing and boundary conditions are slightly different.  In the 1D
case, the evolution was performed with an isometry condition across
the throat.  Consequently, the maximal slicing condition was
implemented with a symmetric boundary condition there.  The 3D
calculation was performed {\em without} an isometry at the throat.
Furthermore, the maximal slicing condition becomes a simple ordinary
differential equation in 1D, whereas in 3D it was solved as a full 3D
elliptic equation with boundary conditions applied on the faces of a
cube.  For these reasons the slicing and geometric meaning of the
coordinates will differ somewhat from the 1D case.

A more geometrically meaningful measurement is of the {\em mass} of
the apparent horizon as defined by Eq.~(\ref{ahmass}), which we
compute in each of four directions, as described in the geodesic
slicing case above.  Results are plotted in Fig.~\ref{fig:maxahmass},
and compared with results obtained for the 1D simulation.  As this is
a Schwarzschild spacetime, $M_{AH}$ should be the Schwarzschild mass
and constant in time.  From the figure one can see that the error is
slightly larger along the diagonal than in the other cases after about
$t=10-15M$.  This can be understood by considering that the effective
resolution in the radial direction is less along the diagonal due to
geometric effects.  The measurements made along the three axes agree
with each other, as expected, and agree reasonably well with measurements
made along the diagonal line. Also, the measurements made along all
four lines agree well with the 1D result, although at the end of the
calculation, when the peaks in metric functions are growing
dramatically (see, e.g., Fig.~\ref{fig:maxgrr}), the 1D result is
slightly better.

As noted above, these results are typical of what is achieved at
``medium'' resolution.  With this resolution, instabilities develop in
the region of strong metric peaks and the code crashes by about
$t=30M$.  At the highest resolution achievable with our present code
and the present NCSA CM-5 (about $180^3$ zones with maximal slicing),
we can reach about $t=35M$.  As the code is second order accurate, the
results are better at higher resolution.

To demonstrate that our code can also evolve a black hole in full 3D,
without symmetries used at the boundaries, we show results for a
maximally sliced black hole placed at the center of the computational
grid.  Because this calculation requires eight times the memory and
computer time, it cannot be run at the same resolution as the
calculation discussed above.  In Fig.~\ref{fig:maxgrrfull} we show a
2D slice through the $z=0$ plane of the metric function $g_{rr}$ on
the full grid at a time $t=15M$.  The spherical nature of this
function is evident.  In Fig.~\ref{fig:maxgrrcompare} we show a
comparison of a 1D cross section of this slice with the same function
obtained by evolving a single octant of the spacetime at the same
resolution with the same outer boundary location.  In the plot, it is
clear that the functions agree extremely well. It turns out, however,
that for the maximal slicing case the results are not {\em exactly}
the same because of the iterative solver used to compute the lapse
function. However, the resulting difference is negligible.

It is clear that maximal slicing presents a number of problems for 3D
black hole evolution.  Not only is it very time (and memory) consuming
to solve a 3D elliptic equation on every time slice, but it is also
difficult to resolve on a 3D Cartesian grid the kinds of very sharp,
spherical step function features that develop in the solution.
Tolerances on the solver must be set quite tightly, especially at late
times, in order to get a good solution.  Towards the end of the
calculation, just before instabilities in the hyperbolic solution set
in, the solution for $\alpha$ can actually become slightly negative at
the ``base'' of the step function, and the derivatives of $\alpha$ can
become unsmooth, both due to difficulties associated with solving for
this function on a 3D Cartesian grid.  These problems can be delayed
by increasing tolerances or resolution, but at some point they will
develop in all cases studied.

\subsection{Algebraic Slicings}
\label{sec:algeresults}

As shown above, maximal slicing is effective for avoiding a black hole
singularity, but in Cartesian coordinates it can develop problems at
late times, and it is very time consuming. Algebraic slicings are cost
effective alternatives to maximal slicing that have not been explored
extensively in numerical relativity.  A theoretical discussion of
slicings we have considered was given in section~\ref{sec:algebraic}
above.  In this section we discuss results obtained for the condition
\begin{equation}
\alpha = f(x^a) (1 + \text{log}(\gamma)),
\end{equation}
where $f(x^a)$ is an arbitrary function of the spatial coordinates.
For more detailed discussion of its properties please refer to
section~\ref{sec:algebraic}.

As the function $f(x^a)$ can be chosen at will, this slicing condition
still has considerable freedom.  The grid stretching problems
associated with black hole evolutions are severe enough in 3D to halt
the evolution due to the unbounded growth in metric functions, as
shown above for maximal slicing.  Therefore, in choosing $f(x^a)$ one
would like to delay the grid stretching effects while allowing true
dynamics to evolve.  A good choice for this is a lapse condition that
vanishes on the black hole throat.  In this case, we choose the
function $f(x^a)$ to have the value of the Schwarzschild lapse and
evolve $\alpha$ with Eq.~(\ref{evolvelapse}). In principle this choice
of slicing condition makes the time coordinate $t$ the Killing time,
so that if the system were evolved analytically the metric would be
truly static.  However, this requires perfect cancellation between
second derivatives of the lapse and Ricci tensor components in the
extrinsic curvature evolution equations (see Eq.~(\ref{excurv
evolution})), which will not occur due to discretization error.
Therefore, the black hole will evolve in time, although the initial
grid stretching effects will be slowed dramatically due to the lapse
profile.  At later times the lapse collapses and develops a profile
similar to the maximal slicing discussed above. Such ``antisymmetric''
(in the appropriate sense across the throat) lapse conditions have
been used extensively in other calculations, such as the collision of
two black holes~\cite{Anninos93b,Anninos94b} and rotating black
holes~\cite{Brandt94b,Brandt94c}.

In Figs.~\ref{fig:alglapse}a,b,c we show the lapse at times $t=0$,
$t=33M$ (partially collapsed) and $t=48M$ respectively for a
simulation run with $128^3$ zones with $\Delta x = 0.06M$.  As the
lapse is antisymmetric across the throat, it approaches the value
$\alpha = -1$
at the origin, and its gradient becomes undefined there.  For this
reason it is essential to perform the evolution with the isometry
condition imposed so that the throat interior is obtained by mapping
the exterior solution rather than through evolution equations.  By time $t=30M$
the lapse has begun to collapse around the throat and by $t=48M$ it
has collapsed dramatically with a profile very similar to that seen
with maximal slicing.

In Fig.~\ref{fig:alggrr} the metric function $g_{rr} =
\gamma_{rr}/\psi^4$ is shown, also at $t=48M$.  As before, this
function has been constructed from the Cartesian metric coefficients,
and shows the familiar peak surrounding the black hole throat.  The
prominent peak structure {\em inside} the throat results from the
mapping of the exterior to the interior region.  As in the previous
cases, we have also compared the evolution with the black hole placed
in the center of the grid to the evolution with the hole in the
corner, with the same results.

Because of the growth of the large gradients in the metric functions,
the evolution becomes unstable shortly after this time, causing the
code to crash by $t=50M$ at this resolution.  Higher resolution can
stall the development of this instability somewhat, but at some point
it develops for all computational parameters tested to date (e.g.,
artificial viscosity parameters, resolution, outer boundary, slicing
choice, etc.).  As we discuss in the next section, the use of a shift
and an apparent horizon boundary condition is a promising way of
avoiding this problem.

\subsection{Apparent Horizon Shift}
\label{sec:ahresults}

It is clear from the results in the above sections that in using
singularity avoiding slicings, independent of the choice of lapse, a
sharp peak will develop in a region slightly inside the horizon, where
the lapse has not completely collapsed.  This is the major limitation
to an accurate long term evolution.  As discussed in
Sec.~\ref{sec:shift} the development of such a peak can be suppressed
by using an apparent horizon boundary
condition~\cite{Seidel92a,Anninos94e}. Here we report on the first
results obtained in testing this condition in 3D.

In this first trial implementation of the apparent horizon boundary
condition, we compute the ``distance freezing'' shift for the 3D
Cartesian evolution by first going back to spherical coordinates,
determining the appropriate shift at each time step and then
transforming the resulting shift vector back to 3D Cartesian
coordinates. The determination of this shift in the spherical
coordinate case has been discussed in detail in ~\cite{Anninos94e}.

Figs.~\ref{fig:ahshiftlapse} and~\ref{fig:ahshiftgrr} show the results
for the lapse function and conformal radial metric function
$g_{rr}=\gamma_{rr}/\psi^4$ (reconstructed from the Cartesian
components) at various times, up to $t=30M$. These results are run
using a $140^3$ grid with cell sizes of $\Delta x=0.15M$ placing the
outer boundary at $\approx 21M$ in each direction.  The data are
displayed along the diagonal line running from the center of the black
hole to the furthest corner.  In this case, the shift vector is
imported and translated from a 1D simulation, but all other
computations in the evolution are performed in the full 3D code.  The
equations are evolved in their most general 3D form, without using
explicitly any simplification due to the particular gauge choice.  In
this simulation, we allow the spacetime to evolve for a short time
($\sim 1M$) before phasing in the shift vector.  During this period,
the maximal slicing condition makes the lapse collapse slightly (as
shown in Fig. \ref{fig:ahshiftlapse}) and the coordinate position of
the horizon moves outward. This motion of the horizon allows for a
``buffer'' region of zones inside the horizon
{}~\cite{Seidel92a,Anninos94e}. From $t\sim 1M$, while the horizon
continues to move outward, we smoothly phase in the distance freezing
shift vector. By $t\sim 2M$ the shift vector is fully phased in and
the horizon remains approximately locked in place at $r \approx 1.5M$.
At this point we drop the inner part ($ r < 1.2M$) of the grid from
the dynamical evolution.  At the inner most grid point retained in the
evolution, we import the values of the metric functions determined
with the spherical code, in the same way as we import the shift
vector.  With the part of the grid which is going to run into the
singularity dropped from the evolution, there is no need to further
collapse the lapse, which is hence held constant in time from this
point onward.  Notice that with such a time independent lapse, which
is non-zero everywhere in the evolved domain, it is {\em not} possible
to evolve the spacetime in the usual treatment without a horizon
boundary condition. The metric functions (of which $\gamma_{rr}$ in
Fig. \ref{fig:ahshiftgrr} is typical) evolve rapidly before and during
the phase-in period, but settle down afterwards.  There is {\it no }
sharp peak, and $\gamma_{rr}$ is of order one throughout the
evolution.  This is to be compared to $\gamma_{rr}$ in
Figs.~\ref{fig:maxgrr} and~\ref{fig:alggrr} above obtained without the
distance freezing shift.  Without a sharp peak in the metric function,
the requirement on resolution is reduced substantially.

Ideally the radial metric function $\gamma_{rr}$ would remain constant
in time with the imported shift vector.  However, at late times there
is a slow downward drifting of the metric functions observed away from
the values at which they are supposed to be ``locked'' (for details,
see~\cite{Seidel92a,Anninos94e}).  We note that, in this
implementation, once the metric functions start drifting, there is
nothing to stop them from drifting further, as the shift is taken to
be a constant in time in this implementation.  What is noteworthy is
that such a simple implementation is already effective.  It nearly
freezes $\gamma_{rr}$ as designed for quite some time, and, more
importantly, the steep peaks observed in the previous sections have
been eliminated.  In comparison to those fast growing sharp peaks, the
slow and rather smooth drift shown in Fig. \ref{fig:ahshiftgrr} should
be considered very satisfactory.

The elimination of steep peaks is important, and is expected to lead
to a much more accurate evolution.  As a check on the accuracy of this
simulation we have computed the location and mass of the apparent
horizon as in the previous sections.  In Fig.~\ref{fig:ahshiftah} we
show the location of the apparent horizon computed during this
simulation as a solid line, and the same quantity computed with a 1D
code at the same resolution without the use of a horizon locking
shift.  After the initial phase-in period, the horizon is firmly
locked in place by the shift, while it continues to move away from the
throat without the shift.  In Fig.~\ref{fig:ahshiftahmass} we show the
apparent horizon mass obtained in the 3D code with the horizon locking
shift, compared to the 1D case without such a shift.  In this first 3D
test case the error in the horizon mass is about the same as the 1D
case without shift at $t=30M$, the final time computed in the 3D case.
However, and more importantly, the slopes of the curves suggest that
the 3D result will be significantly more accurate at later times than
the 1D result without a horizon locking shift.  We consider these
results very satisfactory in this first test of the horizon boundary
condition in 3D, using a highly simplified treatment.  These results
support the claim that apparent horizon boundary conditions are
realizable in the near future and we are currently working on the full
scale horizon boundary treatment.  Progress on that will be reported
elsewhere.

\section{Summary and Future Directions}
\label{sec:summary}

We have developed a general 3D Cartesian code for solving the Einstein
equations in the absence of symmetries.  This code has been applied to
the problem of black hole spacetimes, and we have reported on the
first long term evolution of a black hole in 3D.  Black hole
spacetimes are made quite difficult to study by the need to avoid the
singularity inside the horizon.  We concentrated in this paper
on the evolution of a spherical black hole, since it has the most
troublesome feature (the singularity) and can be studied very
carefully in 1D as a benchmark for what is required of a 3D evolution.
However, because we treat the spherical test problem in a general way
in Cartesian coordinates, we have been able to learn a great deal
about the generic 3D black hole spacetime problem, which we summarize
here.

The boundary conditions on the spacetime can be treated in a number of
ways.  We have demonstrated that the isometry that is commonly applied
at the black hole throat can be applied effectively in a 3D black hole
evolution in Cartesian coordinates, but this is not necessary.  For
some slicing conditions the isometry condition is important in
maintaining accuracy near the throat, but for others it is not
required.  For example, in our maximal slicing simulations, the
isometry is not needed and the entire black hole spacetime inside and
outside the throat can be evolved.

The boundary conditions at the outer edge of the grid are more
delicate, however.  In previous calculations carried out in 1D and 2D,
the outer boundary has been placed far enough away that the metric
could be held static, but in 3D this does not work as well.  We have
tested several conditions that use extrapolation, but in most cases,
extrapolation boundary conditions give worse results than keeping the
outer boundary static. The reason why these methods do not work well
is clear: These methods deal only with the spatial part of the metric,
without taking the time slicing into account.  The time slicing
introduces two effects.  First as the constant $t$ slices, with $t$
not the Killing time of the Schwarzschild geometry (e.g., maximal
slicing with certain boundary conditions), are ``tilted'' with respect
to the Killing one. With the grid points moving normal to the time
slicing as in the case without a shift, they are moving (for maximal
slicing, infalling) with respect to the geometry.  This directly
affects the angular part of the metric functions, here in Cartesian
coordinates some combination of $\gamma_{ab}$.  Secondly, as the
constant $t$ slices are also ``curved'' with respect to the Killing
one, the ``speeds'' with which the grid points move with respect to
the geometry will be different.  This affects the radial part of the
metric function, here a different combination of $\gamma_{ab}$.  With
the assumption that the spacetime geometry near the outer boundary is
locally the same as that of Schwarzschild, both of these two effects
can be taken into account by analyzing how a time slicing, as specified
by a given lapse function, is locally embedded in the Schwarzschild
geometry.  In the present case, as we are actually evolving a
Schwarzschild hole, this boundary condition is exact and can be put
arbitrarily close to the hole.  For a general spacetime, whenever the
outer boundary can be placed far out enough so that the spacetime
there can be approximated by the Schwarzschild geometry locally, this
will provide an accurate outer boundary condition.  We have
successfully constructed such an outer boundary scheme in the 1-D
case, and its extension to 3D is at present under development, and
will be reported elsewhere.

Many slicing and gauge conditions have been tested and reported in
this paper.  For geodesic slicing, we demonstrated that the 3D code
reproduces the results from a 1D code with a high degree of accuracy,
both by comparing metric functions and by looking at derived
quantities such as apparent horizon locations and masses.  A number of
singularity avoiding slicings have been developed for 3D, including
maximal and a class of algebraic slicings.  Maximal slicing works
well, as in 1D and 2D simulations, until the gradients in the metric
become very pronounced.  At that time very tight tolerances on the
elliptic solver are required to improve the accuracy and smoothness of
the lapse solution, but problems develop at late times in all cases
studied.  We have shown that the algebraic slicings are quite
promising and economical, and in some cases we have been able to
evolve beyond $t=50M$.  We regard the algebraic slicings as a major
step forward for 3D black hole evolution, but in {\em all} slicings
studied the evolution cannot be carried beyond a certain point due to
extreme grid stretching effects.  These difficulties have been long
recognized in 1D and 2D studies, but they are more severe in 3D. By
adding more resolution or viscosity terms one can delay the growth of
instabilities, but a more fundamental approach to the problem is
needed.  Different formulations of the equations, such as those
discussed in ~\cite{Bona94b}, allow for the use of other numerical
techniques that may be able to handle the peaks that show up in the
metric functions better.

In order to solve these problems, apparent horizon boundary conditions
are under development by a number of researchers
{}~\cite{Seidel92a,Anninos94e,Thornburg93,Scheel94,Laguna94,Abrahams94b}.
We have demonstrated that a shift vector designed to prevent
coordinates from falling into the hole, combined with cutting away the
singular region inside, can work quite well in a 3D Cartesian black
hole simulation.  In fact, our simulations indicate that with apparent
horizon boundary conditions, 3D calculations can be as accurate or
better than standard 1D calculations without such a boundary
condition.  We are presently working to develop a full implementation
of an apparent horizon boundary condition in 3D.

In the near future we plan to use our code to solve the problems of
the axisymmetric distorted black hole, both rotating and non-rotating,
and also the axisymmetric Misner data for two colliding black holes.
These will be the first truly dynamic black hole spacetimes with
gravitational radiation to be studied in 3D, but they can be compared
with results obtained with mature, axisymmetric codes.  These are all
steps towards the simulation of general, 3D binary black hole
interactions.

Scientific animations of some of our simulations have been prepared
and are kept on our WWW server.  They may be viewed at the URL
http://jean-luc.ncsa.uiuc.edu.

\acknowledgements

We are pleased to acknowledge helpful conversations with a number of
our colleagues, especially Robert Ferrell and Larry Smarr.  The
calculations were performed on the Cray C90 at the Pittsburgh
Supercomputing Center and the Thinking Machines Corporation CM-5 at
NCSA.  This work has been supported by NCSA, by NSF grants Nos.
PHY91-16682, PHY94-07882, PHY94-04788, and PHY/ASC93-18152 (arpa
supplemented), by NASA grant No. NAG 5-2201, and by the joint NSF-NASA
JNNIE project ASC 89-02829.  J.M. acknowledges a Fellowship (P.F.P.I.)
from Ministerio de Educaci\'on y Ciencia of Spain.

\appendix
\section{Code Performance and Programming Strategy for Parallel
Machines}
\label{sec:appendix}

Here we discuss various issues involved in developing our 3D code,
making it efficient for a wide variety of architectures, and
particular numerical issues relevant to the black hole problem.

\subsection{General Code Strategies}
\subsubsection{Portability}
We are currently running our codes on the Thinking Machines CM-5
massively parallel distributed memory system, the Cray C90 vector
multiprocessor and the Silicon Graphics Power Challenge
multiprocessor.  We are extending our codes to run on many other
parallel systems.  In the approach we have taken, the basic
programming language is Fortran~90, a data parallel model.  Extensions
available in High-Performance Fortran (HPF) are used when needed, but
special functions unique to a particular machine are avoided whenever
possible.  In this way the vast majority of the code can be used on
all machines.

To deal with differences among the machines we implement a
preprocessing stage to all the source files.  By using compiler
conditionals to select appropriate code for specific operations, we
are able to maintain a single set of source files that can be used for
compilations on all the systems.  For example, at present, only the
CM-5 supports Fortran~90 array intrinsic functions such as {\tt
MAXLOC(A)}.  In these situations, we use such functions when
supported, but also provide alternate code for other machines to
perform the same operation where necessary.  This has led to a rather
sophisticated build process, but provides great benefits in the
maintenance and development of the code.

\subsubsection{Memory Requirements}
In numerical relativity a large number of variables must be stored.
In the standard ADM split, the basic variables are six metric
functions, six extrinsic curvature variables, three shift components,
and the lapse.  During an evolution, most of the quantities must be
stored on two time slices, requiring approximately 30 variables just
for the evolution alone.  In addition, in order to reduce the
complexity of expressions and to save repeated computation, it is
convenient to compute the contravariant metric components $g^{ab}$,
3-Ricci components $R_{ab}$, and numerous temporary variables
throughout the code, resulting in excess of 50 variables that need to
be stored on every time slice.

However, there are other important considerations for distributed
memory machines.  There are dozens of spatial derivatives that must be
computed on every time slice, and these derivatives appear repeatedly
in the Einstein equations.  Each derivative operation requires
communication between memory locations, which can be computationally
expensive on a distributed memory machine because the data required
for a derivative operation may reside on different processors.  The
communications overhead is more serious for higher order accurate
derivatives (e.g. second vs. fourth order) because more data points
and hence more communications are required.  Therefore, we adopt the
strategy of computing and storing all spatial derivatives on each time
slice, so we may compute derivatives once and reuse the results as
needed.

This technique improves performance dramatically on distributed memory
machines, (although it is unimportant on shared memory machines such
as a Cray C90), but it requires much more storage on each time slice.
However, in 3D if one doubles the storage requirements, the maximum
resolution one can achieve in each direction only decreases by
$2^{(1/3)}$, or about 28\%.  With techniques like these we have been
able to achieve performance for the ``G'' code used in this paper of
nearly 12 Gflops for very large problems on a 512 node CM-5, although
smaller problems run less efficiently.  On that machine, with 16
GBytes of memory, we can perform simulations of about $200^3$ zones.

\subsection{Hyperbolic solvers}
We use an explicit hyperbolic scheme to evolve the spacetime.  Such
algorithms are especially suited to parallel computing, because most
computation is local to a processor.  The spacetime data are
distributed across the various processors, and the evolution of data
on a particular point on the grid depends only on ``nearby''
information in a hyperbolic system.  Communication time between
processors, which is usually the bottleneck in parallel computing, is
required only when computing spatial derivatives via finite
differences that connect different processors.  Communications are
done only once on each time slice, and then extremely long numerical
calculations are performed to evaluate expressions in the evolution
equations.  Therefore, the relevant communications to computation
ratio is very small, leading to excellent performance.  The Einstein
equations are ideal for distributed, parallel computing.

\subsection{Elliptic Solvers}
A difficult code optimization problem comes in solving the elliptic
equations necessary when we evolve the system using maximal slicing.
The code then needs to solve a new elliptic equation on every time
step, although one can save work by solving the equation less
frequently, as it is only a gauge condition.

For efficiency in memory use, the matrix of finite difference
coefficients can be stored as nineteen 3D arrays requiring us to store
only $19N$ elements as opposed $N^{2}$ elements for the full matrix
which consists primarily of zeroes.  This also allows us to store the
coefficients in a manner that relates logically to the computational
grid.  If the elliptic operator for the equation is symmetric in form,
we could also realize further memory conservation by storing only the
upper or lower triangular portions of the coefficient matrix and need
to have only ten 3D arrays. However, due to the communications expense
in implementing this scheme on parallel architectures, it is much more
efficient to store the entire matrix, even when it is symmetric.

When we use maximal slicing to foliate the spacetime, the lapse
function develops a nearly step-function profile, making it difficult
to compute accurate finite difference derivatives of this
function for the hyperbolic evolution part of the code.  Therefore, we
need to solve the maximal slicing equation to a very high tolerance to
resolve sharp features accurately, requiring many iterations of the
solver. Each iteration of the iterative solver requires one or more products
of the sparse coefficient matrix with a vector.  When running the code
on the CM-5, the communications become very time consuming, because in order
to compute the matrix-vector product, each processing node requires
some data from another processor.  Optimization of this communication
is crucial for an efficient implementation.

Investigation into alternative solvers has shown that, particularly
for massively parallel architectures, iterative methods are typically
more efficient than direct solvers.  Recent work in parallel direct
solvers may change this~\cite{Demmel93}, but at present iterative
methods are quite good for problems involving regular
computational meshes.  Experience in development of NCSA 2D
evolution codes indicates that multigrid solvers will play an important
role for very large meshes that will be needed to solve the 3D
evolution problem accurately.  We are currently investigating a
parallel implementation of multigrid for our problem.

\subsection{Code Performance}

The current performance of the code is nearly 12 GFLOPS on the 512
node CM-5 at NCSA. This is achieved using a grid of $200^3$ points and
evolving the system with geodesic slicing. It takes approximately 2 seconds
of CPU time per time step, with every grid point requiring more than
4000 floating point operations per time step.  Algebraic slicings
require a bit more time per iteration due to the use of the isometry
conditions. Typical daily runs of $128^3$ grid points on a 256 node
CM-5 require about 1.5 hours to complete an evolution of 2000 iterations.

With maximal slicing, the elliptic solver slows down the performance
to nearly 6 GFLOPS for a maximum grid size of $180^3$. Both the
performance and the run time strongly depend on the tolerance
imposed on the iterative solver and on how often one solves for the
lapse. Typical runs of $128^3$ take more than 3 hours.

These numbers are constantly changing, as we continue to optimize the code
and new versions of the compiler are released. A complete table
showing the latest performance numbers for the current version of the
code and analysis of linear speedup can be found in our Web server at
http://jean-luc.ncsa.uiuc.edu.


\begin{figure}
\caption
{We show the banded structure of the matrix which results from finite
differencing the maximal slicing equation for the lapse function.
Note that this matrix is sparse and structured, making it suitable for
iterative methods for solution. It is a symmetric matrix with 19
diagonal bands.
\label{fig:19diag}
}
\end{figure}

\begin{figure}
\caption
{We show the radial metric function $A=\gamma_{rr}/\psi^4$ for a
spherical black hole evolved with a 1D code with maximal slicing.
Time slices are shown at intervals of $t=10M$ until the final time of
$t=50M$.  The spacetime shown here was evolved with the logarithmic
$\eta$ coordinate described in the text, with a resolution of $\Delta
\eta = 0.03$.
\label{fig:1dgrr}
}
\end{figure}

\begin{figure}
\caption
{We show the Cartesian coordinate grid with the black hole throat
superimposed as a dark solid line at the lower left corner.  The
throat is located at $r=0.5M$, where $r$ is the isotropic radius, and
the resolution of the grid is $\Delta x = 0.15M$.
\label{fig:grid}
}
\end{figure}

\begin{figure}
\caption
{We compare the radial metric function $A$ obtained with a 1D code
using maximal slicing (solid line) and the ``1+log'' algebraic slicing
(dashed line) discussed in the text.  In these simulations we chose
$\alpha(t=0,r)=1$.  The profiles are shown at intervals $t=10M$.  Both
cases were evolved with $\Delta r = 0.05M$, and develop similar
profiles as the evolution continues.
\label{fig:a1dmaxalg}
}
\end{figure}

\begin{figure}
\caption
{We plot the time required for a time slice to hit the singularity as
a function of resolution for geodesic slicing, computed with our 3D
code.  The analytic result is plotted as a solid line marked by
diamond symbols, while two calculations performed at different
resolutions are plotted with square and circle markers.  As the
resolution is increased, the analytic result is approached.
\label{fig:crashtimes}
}
\end{figure}

\begin{figure}
\caption
{We show a comparison of $\gamma_{rr}/\psi^4$ obtained with 1D and 3D
codes using geodesic slicing. The 1D data were obtained with 128
radial zones and $\Delta r=0.05M$. The 3D data were obtained with
$128^3$ Cartesian grid zones with $\Delta x=\Delta y=\Delta z=0.05M$.
In 1D we evolve $\gamma_{rr}$ while in 3D we reconstruct $\gamma_{rr}$
from the 3D Cartesian metric functions.
\label{fig:geodcompa}
}
\end{figure}

\begin{figure}
\caption
{As in Fig. 6, except that we plot $\gamma_{\theta \theta}/(r^2 \psi^4)$.
\label{fig:geodcompb}
}
\end{figure}

\begin{figure}
\caption
{The metric function $g_{xx}=\gamma_{xx}/\psi^4$ obtained with
geodesic slicing is shown at time $t=3M$. The data are from the $z=0$
plane, and were obtained using $128^3$ grid zones with a
resolution of $\Delta x = 0.05M$.
\label{fig:geogxx}
}
\end{figure}

\begin{figure}
\caption
{The metric function $g_{rr}=\gamma_{rr}/\psi^4$ obtained with
geodesic slicing is shown at time $t=3M$. The data are from the $z=0$
plane. The data were obtained using $128^3$ grid points with a
resolution of $\Delta x = 0.075M$. Note that in this computation the
black hole was placed at the center of the grid.
\label{fig:geofullgrr}
}
\end{figure}

\begin{figure}
\caption
{We plot the position of the apparent horizon for 1D and 3D runs
with geodesic slicing. The 1D data were obtained using 128 grid points
and resolution $\Delta r = 0.0375M$. The 3D data were obtained using
$128^3$ grid points and resolution $\Delta x = 0.0375M$ using a
procedure described in the text.  In all five lines are plotted, but
they are virtually indistinguishable in this figure.
\label{fig:geodah}
}
\end{figure}

\begin{figure}
\caption
{We show the mass of the apparent horizon for 1D and 3D runs with
geodesic slicing. The 1D data were obtained using 128 radial points
and resolution $\Delta r = 0.0375M$. The 3D data were obtained using
$128^3$ grid points and resolution $\Delta x = 0.0375M$.  All results
agree with the analytic result to better than 0.07\% for the entire
evolution.
\label{fig:geodahmass}
}
\end{figure}

\begin{figure}
\caption
{A 2D slice through the plane $z=0$ is shown for the lapse function
$\alpha$ at time $t=28M$ for maximal slicing.  The resolution
is $\Delta x = 0.1M$, and total number of zones is $130^3$.  The singularity
avoiding properties of the maximal lapse create a steep 3D well in $\alpha$
surrounding the throat.
\label{fig:maxalpha}
}
\end{figure}

\begin{figure}
\caption
{A 2D slice through the plane $z=0$ for the metric function
$g_{xx}$ is shown at time $t=28M$ for maximal slicing.  The resolution
is $\Delta x = 0.1M$ and the total number of zones is $130^3$.  The sharp
peak developing in this 3D calculation cannot be resolved, causing
difficulty with the calculation at late times.
\label{fig:maxgxx}
}
\end{figure}

\begin{figure}
\caption
{A 2D slice through the plane $x=0$ is shown for the metric function
$g_{rr}=\gamma_{rr}/\psi^4$ at time $t=28M$ for maximal slicing.  The
resolution is $\Delta x = 0.1M$, and total number of zones is $130^3$.
The sharp spherical peak develops in this 3D calculation just as in
the 1D and 2D calculations.
\label{fig:maxgrr}
}
\end{figure}

\begin{figure}
\caption
{We plot the position of the apparent horizon for 1D and 3D runs with
maximal slicing. The 1D data were obtained using $130$ grid zones with
a resolution $\Delta r = 0.1M$. The 3D data were obtained using
$130^3$ grid zones with a resolution $\Delta x = 0.1M$.
\label{fig:maxah}
}
\end{figure}

\begin{figure}
\caption
{As in Fig. 15, except that we plot the mass of the apparent horizon.
The 1D results are reproduced to within about 5\% by the end of the
calculation.
\label{fig:maxahmass}
}
\end{figure}

\begin{figure}
\caption
{We show a comparison of the metric function $g_{rr}$ computed for a
black hole placed at the center of the numerical grid with one placed
in the ``corner''.  In both cases, the resolution was $\Delta x =
0.075M$ and maximal slicing was used. In (a), the number of grid
points used was $126^3$. However, in (b), the number of grid points
used was $64^3$, with symmetry boundary conditions across the
coordinate planes, giving the same results as in (a).
\label{fig:maxgrrfull}
}
\end{figure}

\begin{figure}
\caption
{A plot of the cross sectional line of $g_{rr}$ is shown for full grid
and corner hole cases for maximal slicing. Data are taken from the
$x-$axis. Note that for $x>0$, both cases are present, but the data for
the two cases are indistinguishable.
\label{fig:maxgrrcompare}
}
\end{figure}

\begin{figure}
\caption
{We show the evolution of the lapse for the `1+log' algebraic slicing
case discussed in the text. Initially, the lapse is Schwarzschild, as
shown in (a). After a while, the lapse begins to collapse (b) as in
the maximal slicing case. Eventually, the lapse is completely
collapsed (c) and the troubles due to steep gradients that also occur in
maximal slicing cause the code to crash shortly after this time.
\label{fig:alglapse}
}
\end{figure}

\begin{figure}
\caption
{We show the function $\gamma_{rr}/\psi^4$ obtained with algebraic
slicing. The data are taken from the $x=0$ plane. $128^3$ grid points
were used, with a resolution of $\Delta x = 0.06M$. The peak inside
the throat is a result of the isometry condition, which maps data
outside the throat to points inside.
\label{fig:alggrr}
}
\end{figure}

\begin{figure}
\caption
{The lapse function is shown for the apparent horizon boundary condition
test along a 1D diagonal line at selected times between $t=0$ and $\sim 30M$. A
nonvanishing shift vector is slowly phased in over the interval $t
\sim 1$ to $2M$ to lock the apparent horizon at a constant coordinate
position.  Also by $t=5M$, the lapse function is frozen so that the
time slicing ceases to be singularity avoiding. Only the points that
are evolved ($r \ge 1.3$) are displayed.
\label{fig:ahshiftlapse}
}
\end{figure}

\begin{figure}
\caption
{A 1D line of the conformal radial metric component
$\gamma_{rr}/\psi^4$ is shown at selected times between $t=0$ and
$\sim 30M$. A nonvanishing shift vector is slowly phased in over the
interval $t \sim 1$ to $2M$ to lock the apparent horizon at a constant
coordinate position and approximately freeze $\gamma_{rr}$. Only the
points that are evolved ($r \ge 1.3$) are displayed.
\label{fig:ahshiftgrr}
}
\end{figure}

\begin{figure}
\caption
{We plot the position of the apparent horizon for the 3D code using a
horizon locking shift and for a typical 1D run using maximal slicing.
\label{fig:ahshiftah}
}
\end{figure}

\begin{figure}
\caption
{We plot the position of the apparent horizon for the 3D code using a
horizon locking shift and for a typical 1D run using maximal
slicing. Clearly, the full 3D simulation using the horizon locking
shift preserves the mass of the apparent horizon better than a typical
1D simulation.
\label{fig:ahshiftahmass}
}
\end{figure}

\end{document}